\documentclass[sigconf]{acmart}

\copyrightyear{2022}
\acmYear{2022}
\setcopyright{acmcopyright}\acmConference[CIKM '22]{Proceedings of the 31st ACM
International Conference on Information and Knowledge Management}{October
17--21, 2022}{Atlanta, GA, USA}
\acmBooktitle{Proceedings of the 31st ACM International Conference on Information
and Knowledge Management (CIKM '22), October 17--21, 2022, Atlanta, GA, USA}
\acmPrice{15.00}
\acmDOI{10.1145/3511808.3557314}
\acmISBN{978-1-4503-9236-5/22/10}

\usepackage{multirow}
\usepackage{multicol}
\usepackage{subfigure}
\usepackage{bm}
\usepackage{color}
\usepackage[doipre={doi:~}]{uri}

\AtBeginDocument{%
  \providecommand\BibTeX{{%
    \normalfont B\kern-0.5em{\scshape i\kern-0.25em b}\kern-0.8em\TeX}}}

\begin{document}

\title{Evolutionary Preference Learning via Graph Nested GRU ODE \\ for Session-based Recommendation}

\author{Jiayan Guo$^{*}$}\thanks{$^{*}$Both authors contribute equal to this work.}
\email{guojiayan@pku.edu.cn}
\affiliation{\institution{Peking University}\city{Beijing}\country{China}}

\author{Peiyan Zhang$^{*}$}
\email{pzhangao@cse.ust.hk}
\affiliation{\institution{HKUST}\country{Hong Kong}}

\author{Chaozhuo Li}
\email{cli@microsoft.com}
\affiliation{\institution{Microsoft Research Asia}\city{Beijing}\country{China}}

\author{Xing Xie}
\email{xing.xie@microsoft.com}
\affiliation{\institution{Microsoft Research Asia}\city{Beijing}\country{China}}

\author{Yan Zhang}
\email{zhyzhy001@pku.edu.cn}
\affiliation{\institution{Peking University}\city{Beijing}\country{China}}

\author{Sunghun Kim}
\email{hunkim@cse.ust.hk}
\affiliation{\institution{HKUST}\country{Hong Kong}}

\renewcommand{\shortauthors}{Trovato and Tobin, et al.}

\begin{abstract}

Session-based recommendation (SBR) aims to predict the user’s next action based on the ongoing sessions. Recently, there has been an increasing interest in modeling the user preference evolution to capture the fine-grained user interests. While latent user preferences behind the sessions drift continuously over time, most existing approaches still model the temporal session data in discrete state spaces, which are incapable of capturing the fine-grained preference evolution and result in sub-optimal solutions. To this end, we propose Graph Nested GRU ordinary differential equation~(ODE), namely GNG-ODE, a novel continuum model that extends the idea of neural ODEs to continuous-time temporal session graphs. The proposed model preserves the continuous nature of dynamic user preferences, encoding both temporal and structural patterns of item transitions into continuous-time dynamic embeddings. As the existing ODE solvers do not consider graph structure change and thus cannot be directly applied to the dynamic graph, we propose a time alignment technique, called t-Alignment, to align the updating time steps of the temporal session graphs within a batch. Empirical results on three benchmark datasets show that GNG-ODE significantly outperforms other baselines.

 
\end{abstract}

\ccsdesc[500]{Information systems~Recommender systems}

\keywords{Session-based Recommendation; Graph Neural ODE; Graph Neural Networks}

\maketitle

\section{Introduction}

Recommender systems can help provide users with personalized information according to their preferences reflected in the historical interactions~\cite{he2017neural}, which are widely applied in e-commerce websites, web searches, and so forth~\cite{zhang2019deep,wang2021survey}. However, in some scenarios where only the user’s recent interactions within a narrow time range are available, the general recommenders are not applicable since the collaborative signal is scarce, leading to the obscure of user preferences~\cite{Hidasi2016SessionbasedRW}. Thus, session-based recommendation (SBR) is proposed to detect the user intent from the limited interactions in the current session and make recommendations, where the session is defined as the user’s actions within a period of time~\cite{Hidasi2016SessionbasedRW,li2017neural}.

\begin{figure}[t]
    \centering
    \includegraphics[width=.9\linewidth]{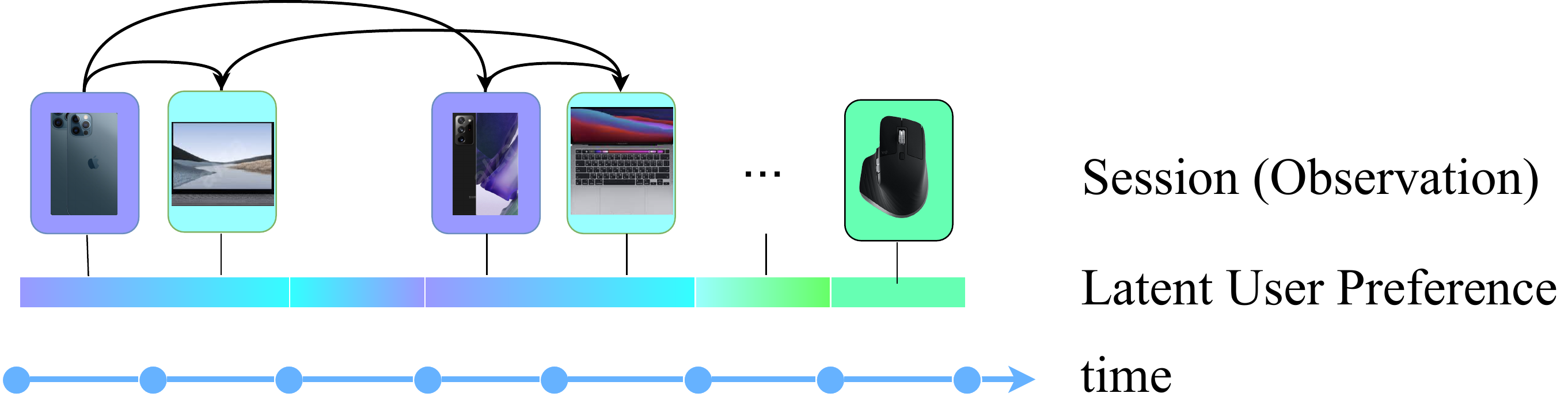}
    \caption{Illustration of the complex structural and temporal patterns in the session. Arrows denote that nearby and distant transition dependencies co-exist. The color gradient  represents the continuity change of user preferences as time progresses. Item clicks can be interpreted as the observations of latent continuous user preferences.}
    \label{fig:sbr_exp}
    \vspace{-0.5cm}
\end{figure}

Most existing SBR methods focus on modeling sequential pattern among items of a session by using Recurrent Neural Networks (RNNs)~\cite{Hidasi2016SessionbasedRW,li2017neural,wang2019collaborative} or Graph Neural Networks (GNNs)~\cite{wu2019session,Qiu2019RethinkingTI,Wang2020GlobalCE,guo2022learning}. However, these works view a session as a short sequence and assume that the primary intention of the user in a session usually remains the same, and try to capture the user’s preference directly from the entire session. Consequently, they often ignore the fact that a user’s fine-grained preference can drift over time, even in a relatively short-term session. Although the temporal pattern is crucial in capturing the fine-grained user preference, research on utilizing temporal information in SBR is still in the early stage.

Fortunately, multiple lines of recent studies in SBR have aimed to embrace this challenge by incorporating additional temporal information.
The first line of works~\cite{pan2021dynamic,zhou2021temporal} models the evolution of user preference in a discrete-time setting. They model a session as snapshots of a dynamic session graph sampled at fixed-length timestamps. Thus, these approaches cannot model the irregularity of time intervals, which is essential for analyzing complex dynamics of user preferences, \textit{e.g.,} when the dwelling time of a user on an item becomes shorter, the user's interest in the item tends to decrease~\cite{fan2021continuous}.
Another line of works integrates the time dimension by considering timestamps information as a contextual dimension~\cite{DBLP:journals/corr/abs-2112-15328,DBLP:conf/sigir/ZhouTHZW21}. However, these methods generate discrete user preferences that ignore the time elapse effect. Consider a user who makes a purchase today and her preference representation is updated. The representation will stay the same regardless of when she returns (\textit{i.e.,} a day later, a month later). As a result, the same recommendations will be offered when she returns next time. However, user preferences may change over time~\cite{cheng2017predicting,kumar2019predicting}. The time elapse effect on user preferences should be considered and thus the preference representation needs to be updated to the query time. 
In this paper, we argue that the user's preference is a continuous concept evolving as time progresses. As shown in Figure~\ref{fig:sbr_exp}, item interactions can be interpreted as the observations of the latent continuous user preference at a specific timestamp. By modeling the preference dynamics in the continuous-time setting, we no longer need the equal-length time slice segmentation of the whole timeline and manage to consider the time elapse effect to predict the future embedding trajectories of items as time progresses.

In particular, modeling the user preference in a fully continuous manner is challenging. As most neural networks are discrete, where the iterative update of hidden states between two layers is a discretization of a continuous transformation~\cite{chen2018neural,lu2018beyond,haber2017stable}, they cannot model user preference in the continuous-time setting. To handle this challenge, in this paper, we propose to utilize neural ODE to complete this task. Owing to its intrinsic continuous nature, neural ODE enables tracking of the evolution of the underlying system. It is expected to offer improved performance compared with using discrete methods to model a continuous dynamical system~\cite{huang2020learning,huang2021coupled}. However, directly applying neural ODE models in SBR is still inapplicable. As user-item interactions occur irregularly along time, the neural ODE model should be theoretically continuous to guarantee stability. Moreover, due to the dynamic nature of a temporal session graph, the updating time steps of the temporal session graph within a batch may not be consistent. Thus, the existing ODE solvers are inapplicable in the batch update process since the solvers could accept only a one-time step parameter for each calculation.

To address these issues, we propose a novel ODE-based model for modeling the dynamics of user preference along time in a fully continuous manner.
Different from previous snapshots-based methods, given an ongoing session, we transform the session into a fully continuous temporal session graph without using snapshots, which builds the potential structural and temporal relations between items. Afterward, we employ Graph Gated Neural Network~(GGNN)~\cite{li2015gated,wu2019session} to encode the item embeddings and transition patterns simultaneously to infer the latent initial states for all items. We further derive a \underline{G}raph \underline{N}ested \underline{G}RU~\cite{li2019predicting,skardinga2021foundations,seo2018structured,pareja2020evolvegcn} inspired continuous-time 
\underline{O}rdinary \underline{D}ifferential \underline{E}quation network (GNG-ODE) that propagates the latent states of the items between different time steps as time progresses. Different from most existing temporal SBR models that
learn the dynamics by employing recurrent model structures with discrete depth, our model coincides the time domain  with the depth of a neural
network and takes advantage of ODEs to steer the latent user/item features between two timestamps smoothly.
As the existing ODE solvers are inapplicable in SBR where the graph is dynamic, we further propose a time alignment algorithm, called \textit{t-Alignment}, to adapt the existing ODE solvers onto our dynamic graph setting by aligning the updating time steps of the dynamic session graphs within a batch.
We conduct extensive experiments on three real-world public benchmarks. Comprehensive experimental results verify that GNG-ODE significantly outperforms the competitive baselines.

Our primary contributions can be summarized as follows:
\begin{itemize}
    \item We propose a novel GNG-ODE to effectively consider the intrinsic complex nature of user-item interactions by modeling a session as a continuous-time temporal session graph. In this carefully designed graph structure, the temporal information of item transitions is preserved in a fully continuous manner. To the best of our knowledge, this is the first work to model the continuous evolution of user preference using neural ODE in SBR.

    \item We show that GNG-ODE is theoretically well-posed, \textit{i.e.,} its solution always exists and is unique (see Section~\ref{sec:proof}). Besides, it enjoys several good numerical properties. We also propose the \textit{t-Alignment} algorithm to make existing ODE solvers applicable to dynamic environments in SBR.
    \item Extensive experiments on three public datasets demonstrate the effectiveness of our GNG-ODE model. Compared with all competitive baselines, the improvements brought by modeling the continuous evolution of user preferences are at most 6.05\%, according to the ranking metric on average.
\end{itemize}
\vspace{-0.15cm}
\section{Related Works}

\textbf{Session-based Recommendation.} Following the development of deep learning, many neural network based approaches have been proposed for SBR. Hidasi et al.~\cite{Hidasi2016SessionbasedRW} first propose to leverage the recurrent neural networks (RNNs) to model users' preferences.
Afterward, attention-based mechanisms are incorporated into the system and significantly boost performance. NARM~\cite{Li2017NeuralAS} utilizes attention on RNN models to enhance the captured features while STAMP~\cite{Liu2018STAMPSA} captures long and short-term preferences relying on a simple attentive model. Convolution Neural Networks (CNNs) are also leveraged. Tang et al.~\cite{tang2018personalized} try to embed item session as a matrix and perform convolution on the matrix to get the representation.

To better model the transitions within the sessions, most recent developments focus on leveraging Graph Neural Networks (GNNs) to extract the relationship between sessions. Wu et al.~\cite{Wu2019SessionbasedRW} first propose to capture the complex transitions with graph structure. Afterward, Pan et al.~\cite{Pan2020StarGN} try to avoid overfitting through highway networks~\cite{Srivastava2015HighwayN}. Position information~\cite{Wang2020PAGGANSR}, target information~\cite{Yu2020TAGNNTA}, and global context~\cite{Wang2020GlobalCE} are also taken into consideration to further improve the performance. \\

\noindent\textbf{Temporal Information in SBR.} Temporal information plays a vital role in user preference modeling. Although there are a few works in other recommendation areas~\cite{li2020time,vassoy2019time,bai2019ctrec,li2020time,kumar2019predicting,fan2021continuous,chen2021learning} utilize the temporal information to facilitate recommendation, the temporal-related method has not been fully explored in SBR. Some prior efforts reduce the temporal information into the relative order/position information. For example, Yu et al.~\cite{yu2016dynamic} use RNN to capture the sequential signal, which reveals the user’s future dynamic preference in the next-basket recommendation. Pan et al.~\cite{pan2021dynamic} further model the evolution of item transitions by constructing a sequence of dynamic graph snapshots which contains the graphs transformed from the session at different timestamps. Zhou et al.~\cite{DBLP:journals/corr/abs-2112-15328,DBLP:conf/sigir/ZhouTHZW21} integrate the time dimension by considering timestamps information as a contextual dimension. The observations of user clicks are put into bins of fixed duration, and the latent dynamics are discretized in the same way. To characterize the dynamics from both the user side and the item side, Zeyuan et al.~\cite{chen2021learning} propose to build a global user-item graph for each time slice and exploit time-sliced graph neural networks to learn user/item embeddings. 


As outlined above, we find previous works on SBR have some limitations.
\textbf{First}, temporal information is rarely or crudely exploited in these
works. \textbf{Second}, existing methods model structural and temporal patterns separately without considering their interactions, which restricts the capacity of the models. \textbf{Third}, some methods rely on the segmentation of the whole timeline into a specified number of equal-length time slices, resulting in the temporal information loss problem~\cite{chen2021learning,zhou2021temporal}. \textbf{Finally}, these methods generate discrete user preference representations that ignore the time elapse effect on user preferences. The representation will stay the same regardless of when the user returns to the platform, \textit{i.e.,} a day later, a week later, or even one month later, thus limiting the performance. \\

\noindent\textbf{Neural Ordinary Differential Equation.} Neural ODE is a continuous approach to model the discrete sequence governed by a time-dependent function $\mathcal{T}\rightarrow\mathbb{R^d}$ of a continuous time variable $t$. 
\begin{equation}\label{diff1}
    \frac{d\bm{h}(t)}{dt}=\bm{f}_\theta(\bm{h}(t), t)
\end{equation}

\noindent where $\theta$ is the parameter of the differential function. Eq.~(\ref{diff1}) drives the system state forward in infinite steps over time. The differential function $f: \mathbb{R}^d\times \mathbb{R}\rightarrow \mathbb{R}^d$ induces a differential ﬁeld that covers the input space. Given initial state $h(t_0)$, we can derive the state of time $T$ by a black-box differential equation solver, which evaluates the hidden unit dynamics $f$ wherever necessary to determine the solution with the desired accuracy.
\begin{equation}\label{diffsolve}
    \bm{h}(T)=\bm{h}(t_0)+\int_{t=t_0}^T\bm{f}_\theta(\bm{h}(t),t)dt
\end{equation}

There is a rich body of literature on Neural ODE recently. Ricky TQ et al.~\cite{chen2018neural} first propose the Neural ODE framework and develop an adjoint method to solve the ode function, which is memory efficient. To improve the expression ability of Neural ODE, Junteng et al.~\cite{DBLP:conf/nips/JiaB19} provide a data-driven approach to learn continuous and discrete dynamic behaviors, Emilien et al.~\cite{DBLP:conf/nips/DupontDT19} add extra dimensions in hidden space and Cagatay et al.~\cite{DBLP:conf/nips/YildizHL19} propose second-order Neural ODE. To enable better learning on irregular sampled sequential data, Yulia et al.~\cite{DBLP:journals/corr/abs-1907-03907} propose to combine RNN and Neural ODE and Edward et al.~\cite{NIPS2019_8957} introduce a RNN-based ODE that uses GRU-Bayes to update the hidden state. Michael et al.~\cite{DBLP:journals/corr/abs-1911-07532} first introduce graph Neural ODE that models the diffusion process on graphs and achieves better results than discrete versions on various tasks. Louis-Pasca et al.~\cite{DBLP:journals/corr/abs-1912-00967} derive the analytic solution to the graph Neural ODE that avoids using of ODE solvers. Chengxi Zang and Fei Wan~\cite{DBLP:journals/corr/abs-1908-06491} introduce the graph Neural ODE on dynamic graphs while do not consider the change of graph structure.  Ziwei et al.~\cite{choi2021lt} first utilize Neural ODE to learn the optimal layer combination of collaborative filtering model rather than relying on manually designed architecture. However, how to exploit Neural ODE to learn the temporal dynamics in SBR still remains unexplored.


\begin{figure}[t]
\subfigure[Static Session Graph]{\centering
    \includegraphics[width=.46\linewidth]{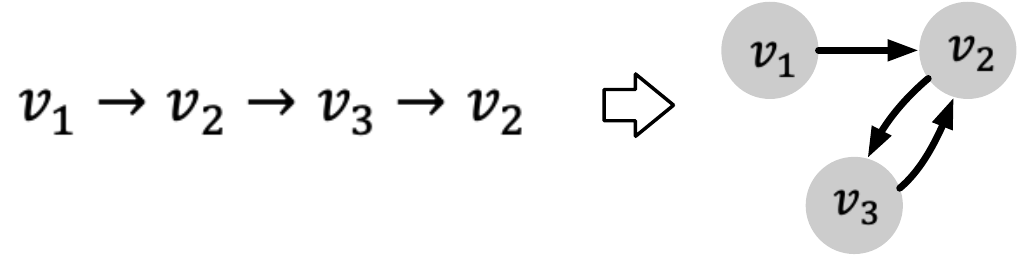}
    \label{fig:session_graph}}
\subfigure[Temporal Session Graph]{\centering
    \includegraphics[width=.46\linewidth]{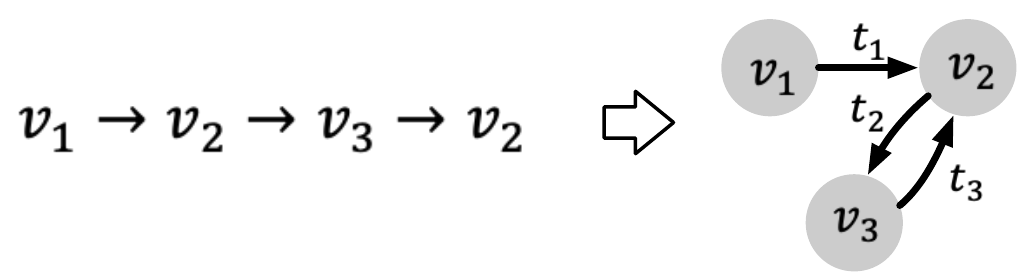}
    \label{fig:temporal_session_graph}}
    \caption{An illustration of (a) session graph and (b) temporal session graph. For (b), when a user clicks an item, the topological structure of the session graph changes.}
    \vspace{-0.5cm}
\end{figure}

\section{Problem Definition}

Assume the item set is $V=\{v_1, v_2,..., v_{|V|}\}$, where $v_{i}$ indicates item $i$ and $|V|$ is the number of all items. Given an ongoing session denoted as $S=\{v_1, v_2,..., v_{n}\}$, the aim of a session-based recommendation is to predict the items that the user will interact with at the next timestamp, that is, $v_{n+1}$. Specifically, the session-based recommender system takes the session $S$ as input and outputs the prediction scores on all candidate items, then the items ranked at the top $K$ positions will be recommended to the user. 

\section{Approach}

\begin{figure}[tbp]
    \centering
    \includegraphics[width=.9\linewidth]{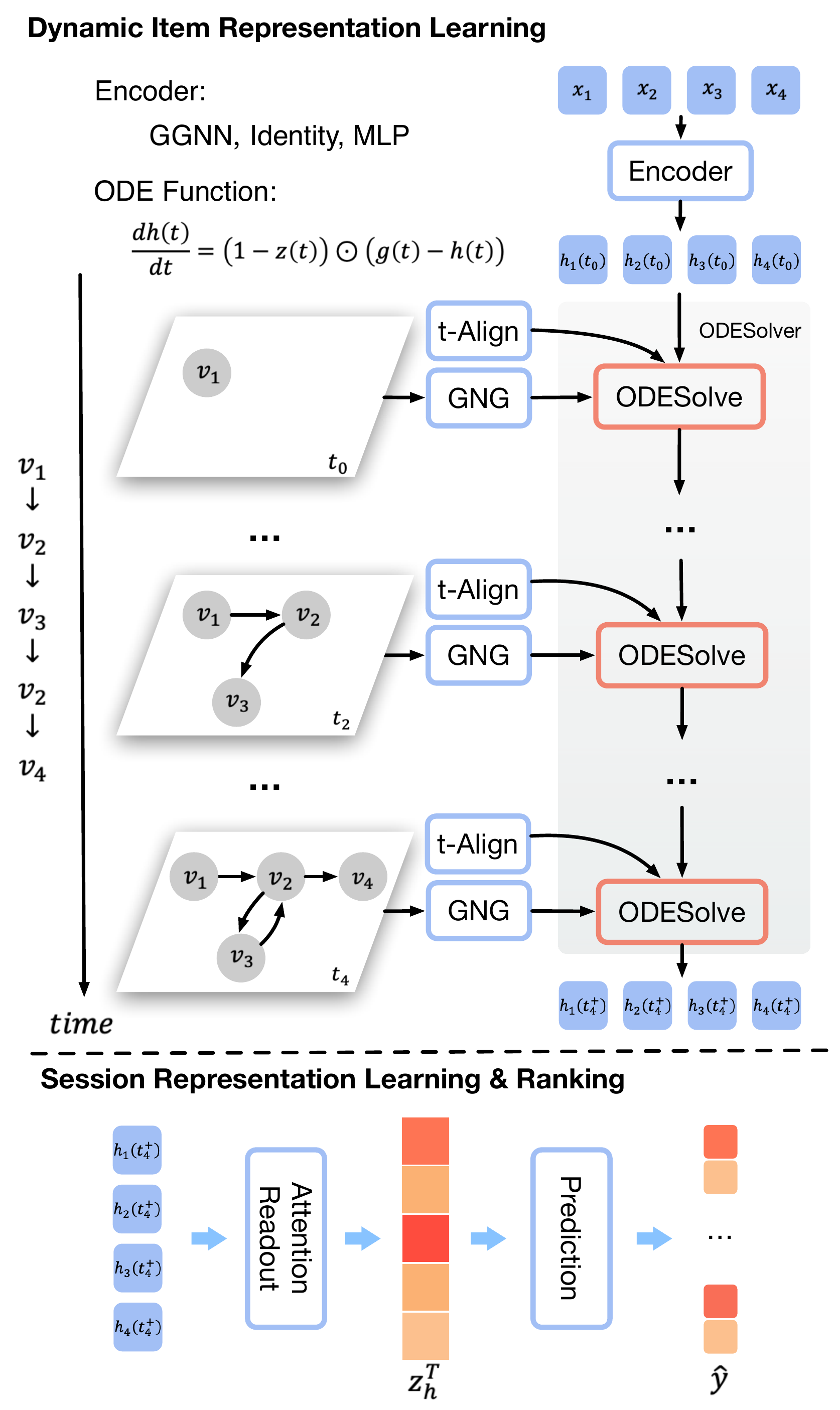}
    \caption{Overview of the GNG-ODE framework. First, the item embeddings are transformed to form the initial hidden states. Then the GNG-ODE is applied to infer the latest item representations. Finally, we apply attention readout to generate session representations to predict the next item.}
    \label{fig:gng_ode}
    \vspace{-0.5cm}
\end{figure}

 In this section, we describe our proposed Graph Nested GRU Ordinary Differential Equation for Session-based Recommendation~(GNG-ODE) in detail, which is constituted of three main components, that is, ~(i) the temporal session graph construction, ~(ii) the dynamic item representation learning and ~(iii) the user preference generation and prediction. The framework of the proposed GNG-ODE is schematically shown in Figure~\ref{fig:gng_ode}. Given an ongoing session, we first construct a temporal session graph which contains the graphs transformed from the sessions at different timestamps. Then, we learn the dynamic item representations through GNG-ODE. Finally, we generate the hybrid user preference, which is utilized to make predictions on candidate items.

\subsection{Temporal Session Graph Construction}
Given a session $S=\{v_1, v_2,..., v_{n}\}$, we first generate the updating time point of the session and its corresponding target items at different timestamps as $(\widetilde{S}_{1}, v_{2}), (\widetilde{S}_{2}, v_{3}),..., (\widetilde{S}_{n-1}, v_{n})$, where $\widetilde{S}_{t}=\{v_1, v_2,..., v_{t}\}$. This is similar to the data augmentation method widely applied in SBR~\cite{wu2019session,pan2020star}. However, different from the existing methods which shuffle the augmented samples and utilize them for training individually, in GNG-ODE we capture the evolution of the session graphs over multi-time steps. Specifically, as shown in Figure~\ref{fig:temporal_session_graph}, we construct a continuous time dynamic session graph denoted as $\mathcal{G}=(\mathcal{V},\mathcal{E}, \tau)$, where $\mathcal{V}=\{1,...,N\}$ is the set of all items appeared in the session. $\mathcal{E}\subseteq\mathcal{V} \times \mathcal{V}$ is the multiset of all edges in the session graph representing the transition between two items. A time  function, $\tau: \mathcal{E}\rightarrow \mathbb{R}^{+}$, maps each transition edge into time space. Under such temporal session graphs, we can learn different item embeddings of the items for various timestamps, which helps generate dynamic item representations for accurate user preference generation, as stated in~\cite{xu2019graph}. 

\subsection{Dynamic Item Representation Learning}
\label{sec:proof}
User click history can be seen as irregularly-sampled data that represents the observations of latent user interests. Typically, observations of user clicks are put into bins of fixed duration~\cite{DBLP:journals/corr/abs-2112-15328,DBLP:conf/sigir/ZhouTHZW21}, and the latent dynamics are discretized in the same way. This leads to difficulties with missing data~(\textit{e.g.,} when there is no clicks at some time points) and ill-defined latent variables~\cite{chen2018neural}. To handle these challenges, our representation learning part consists of three components: ~(i) A GNN-inspired encoder that transforms the transition structure of observed items into initial hidden states. ~(ii) A hidden trajectory prediction model characterized by the GNG-ODE function and t-Alignment technique to learn the latent dynamics of the transition evolution. ~(iii) An attention-based decoder that generates the distribution of the next item that the user may click. 

\subsubsection{\textbf{Initial Latent State Encoder}}

Items in a session can be regarded as observations of the latent user preference. To capture the dynamics of latent user preference, we first transform the raw embeddings of items in a session and their static transitions into the initial latent representations by GGNN. GGNN is widely used in session-based recommendation tasks~\cite{wu2019session,chen2020handling}. Given a static session graph $\mathcal{G}=(\mathcal{V},\mathcal{E})$, the GGNN first aggregates neighborhood information to form a neighborhood representation for a node, then applies GRU to combine the original node representation and the neighborhood representations:

\begin{equation}
\vspace{-0.3cm}
    \begin{split}
        \overline{\bm{h}}_{u}^{(l)}&=\sum_{v\in\mathcal{N}_{u}}\bm{w}_{uv}\bm{h}_{v}^{(l)} \\
        \bm{h}^{(l+1)}_u&=\text{GRU}\left (\bm{h}_u^{(l)}, \overline{\bm{h}}^{(l)}_{u}\right )
    \end{split}
    \label{eq:ggnn}
    \vspace{-0.3cm}
\end{equation}




\noindent where ${\bm{h}}_{u}^{(l)}$ is the hidden representation of item $u$ in layer $l$, while $\overline{\bm{h}}_{u}^{(l)}$ denotes the neighborhood representation of item $u$ in layer $l$. $w_{uv}$ is the edge weight of edge $e_{uv}$ . Details of constructing the static session graph can be found in~\cite{wang2021survey}. By applying GGNN we can infer a initial states jointly considering both item attributes and transition patterns, thus benefits the preference modeling capacity. The output of the last layer is then normalized by $l_2$-norm to make the value between $[-1,1]$ to ensure the stability of ODE solvers. We denote it as $\bm{h}_{u}$ for simplicity.

\subsubsection{\textbf{Graph Nested GRU ODE}}
After computing the latent initial states for items, we now define the Graph Nested continuous-time GRU ODE~(GNG-ODE) function that drives the system to move forward. Graph Nested GRU~(GNG) is widely applied in dynamic graph learning settings~\cite{li2019predicting,skardinga2021foundations,seo2018structured,pareja2020evolvegcn}, by

\begin{equation}
\vspace{-0.1cm}
    \begin{split}
    \bm{r}_u^t &=\sigma\left (W^r_{\mathcal{G}}\bm{x}^t_u+U^{r}_{\mathcal{G}}\bm{h}_{u}^{t-1}+\bm{b}_r\right ) \\ 
    \bm{z}_u^t &=\sigma\left (W^z_{\mathcal{G}}\bm{x}^t_u+U^{z}_{\mathcal{G}}\bm{h}_{u}^{t-1}+\bm{b}_z\right) \\
    \bm{g}_u^t &=\tanh\left (W^h_{\mathcal{G}}\bm{x}^t_u+U^{h}_{\mathcal{G}}(\bm{r}_u^t\odot \bm{h}_u^{t-1})+\bm{b}_h\right) \\
    \bm{h}_u^t &=\bm{z}_u^t\odot \bm{h}_u^{t-1}+(1-\bm{z}_u^t)\odot{\bm{g}_u^t} 
    \label{eq:gng}
    \end{split}
\end{equation}

\noindent where $\bm{r}_u^t$ and $\bm{z}_u^t$  are the reset gate and select gate at time step $t$ respectively. Let $\bm{x}_u^t$ and $\bm{h}_u^t$ denote the input embeddings and hidden state of item $u$ of time step $t$, respectively. $\bm{b}_r,\bm{b}_z,\bm{b}_h\in \mathbb{R}^{d}$ are the parameters and $W_{\mathcal{G}},U_{\mathcal{G}}$ denote one layer of graph convolutional networks~\cite{kipf2016semi} to aggregate neighborhood information of item $u$. GNG models the structural and temporal dependency and performs well on discrete-time dynamic graphs. Here we show how to derive a continuous-time GNG-ODE. Specifically, we firstly show that the form of GNG can be written as a difference equation. Given the standard update for the hidden state $\bm{h}_u^{t}$ of the GNG in Eq.~(\ref{eq:gng}):

\begin{equation}
    \bm{h}_u^t =\bm{z}_u^t\odot \bm{h}_u^{t-\Delta{t}}+\left(1-\bm{z}_u^t\right )\odot{\bm{g}_u^t}
    \label{eq:ode_diffs}
\end{equation}

We can obtain a difference equation by subtracting $\bm{h}_u^{t-\Delta t}$ from this state update equation and factoring out $(1-\bm{z}_u^t)$:
\begin{equation}
    \begin{split}
    \Delta \bm{h}_u^t &= \bm{h}_u^{t}-\bm{h}_u^{t-\Delta t} 
    \\
    &= \left (1-\bm{z}_u^t\right )\odot \left (\bm{g}_u^t-\bm{h}_u^{t-\Delta t}\right )
    \label{eq:odef_diff_pre}
    \end{split}
\end{equation}
This difference equation naturally leads to the following ODE for $\bm{h}(t)$ when $\Delta t\rightarrow 0$:
\begin{equation}
    \begin{split}
    \frac{d\bm{h}_u(t)}{dt}=\left (1-\bm{z}_u(t)\right )\odot \left (\bm{g}_u(t)-\bm{h}_u(t)\right )
    \label{eq:odef_diff}
    \end{split}
\end{equation}

\noindent with $\bm{r}_u(t)$, $\bm{z}_u(t)$, $\bm{g}_u(t)$ the following forms:

\begin{equation}
\centering
    \begin{split}
        \bm{r}_u(t)&=\sigma\left(W_{\mathcal{G}}^r\bm{x}_u(t)+U_{\mathcal{G}}^r \bm{h}_u(t)+\bm{b}_r\right ) \\
        \bm{z}_u(t)&=\sigma\left(W_{\mathcal{G}}^z\bm{x}_u(t)+U_{\mathcal{G}}^z \bm{h}_u(t)+\bm{b}_z\right ) \\
        \bm{g}_u(t)&=\text{tanh}\left(W_{\mathcal{G}}^h\bm{x}_u(t)+U_{\mathcal{G}}^h\left (\bm{r}_u(t)\odot \bm{h}_u(t)\right )+\bm{b}_h\right )
    \end{split}
\end{equation}


The form of derived GNG-ODE is in line with GRU-ODE family. As time intervals between user clicks are irregular, the ODE function should theoretically ensure numerical stability. Inspired by~\cite{NIPS2019_8957}, we have the following corollaries.

\

\noindent\textbf{Corollary 1.} $\bm{h}_u(t)$ is bounded by $[-1, 1]$.

\noindent\textit{Proof.} This bound comes from the negative feedback term in Eq.~(\ref{eq:odef_diff}), which stabilizes the resulting system. In detail, as we use $L_2$-norm to normalize the initial latent state, the $j$-th dimension of the starting state $\bm{h}_u(0)$ is within $[-1, 1]$, then $\bm{h}(t)_j$ will always stay within [-1, 1] because

\begin{equation}
    \frac{d\bm{h}_u(t)_j}{dt}|_{t:\bm{h}_u(t)_j=1}\le 0 \ \ \text{and} \ \ \frac{d\bm{h}_u(t)_j}{dt}|_{t:\bm{h}_u(t)_j=-1}\ge 0
\end{equation}

\noindent This can be derived from the ranges of $\bm{z}$ and $\bm{g}$ in Eq.\ref{eq:gng}. Moreover, when $\bm{h}_u(0)$ start outside of $[-1, 1]$, the negative feedback will quickly push $\bm{h}_u(t)$ into this region, making the system also robust to numerical errors.

\

\noindent\textbf{Corollary 2.} GNG-ODE is Lipschitz continuous with constant $K=2$.

\noindent\textit{Proof.} As $\bm{h}$ is differentiable and continous on $t$, based on mean value theorem~\cite{flett19582742} that for any $t_a,t_b$, there exists $t^*\in(t_a,t_b)$ such that

\begin{equation}
    \bm{h}(t_a)-\bm{h}(t_b)=\frac{d\bm{h}_u(t^*)}{dt}(t_a-t_b)
\end{equation}

\noindent Taking the euclidean norm of the previous expression, we ﬁnd

\begin{equation}
    ||\bm{h}_u(t_a)-\bm{h}_u(t_b)||=||\frac{d\bm{h}_u(t^*)}{dt}(t_a-t_b)||
\end{equation}

\noindent Furthermore, we have shown that $\bm{h}_u(t)$ is bounded on $[-1,1]$. Hence, because of the bounded functions appearing in the ODE (sigmoids and hyperbolic tangents), the derivative of $\bm{h}_u(t)$ is itself bounded by $[-2, 2]$. We conclude that $\bm{h}_u(t)$ is Lipschitz continuous with constant $K=2$.

Based on the above corollaries, our GNG-ODE enjoys the following properties: 

\noindent\textbf{Continuity.} It means that our training procedure is further tractable. Specifically, The Cauchy–Kowalevski theorem~\cite{folland2020introduction} states that, given $\bm{f}=\frac{d\bm{h}_u(t)}{dt}$, there exists a unique solution
of $\bm{h}$ if $\bm{f}$ is analytic (or locally Lipschitz continuous), \textit{i.e.,} the ODE
problem is well-posed if $f$ is analytic. In our case, as the GNG-ODE is Lipschitz continuous with constant $K=2$, there will be only a unique optimal ODE for $\bm{h}_u(t)$. Owing to the uniqueness of the solution, we could find a good solution for GNG-ODE function. Our method becomes fully continuous that can derive item representations at any given timestamps and any time granularity. In this way, we further avoid generating discrete user preference representations and manage to model the time elapse effect to predict the future embedding trajectories of items as time progresses.

\noindent\textbf{Robustness.} The continuous nature of our model allows it to track the evolution of the underlying system from irregular observations, and no longer need the equal-length slice segmentation of the whole timeline, which empowers our method to perceive more fine-grained temporal information compared with previous methods.


We can then apply various ODE solvers to integrate the ODE function in Eq. (\ref{eq:odef_diff}). ODE solvers discretize time variable
$t$ and convert an integral into many steps of additions. Widely used solvers are fix-step solvers like explicit Euler and fourth-order Runge–Kutta~(RK4) method or adaptive step solvers like Dopri5. Then the item representation at time $T$ can be inferred by:

\begin{equation}
    \bm{h}_u(T) = \bm{h}_u(t_0) + \int_{t=t_0}^T\frac{d\bm{h}_u(t)}{dt}
\end{equation}

\noindent where $\bm{h}_u(t_0)$ is the initial hidden state of node $u$ derived by the encoder in Section 3.2.1. 

\begin{figure}[htbp]
    \centering
    \includegraphics[width=\linewidth]{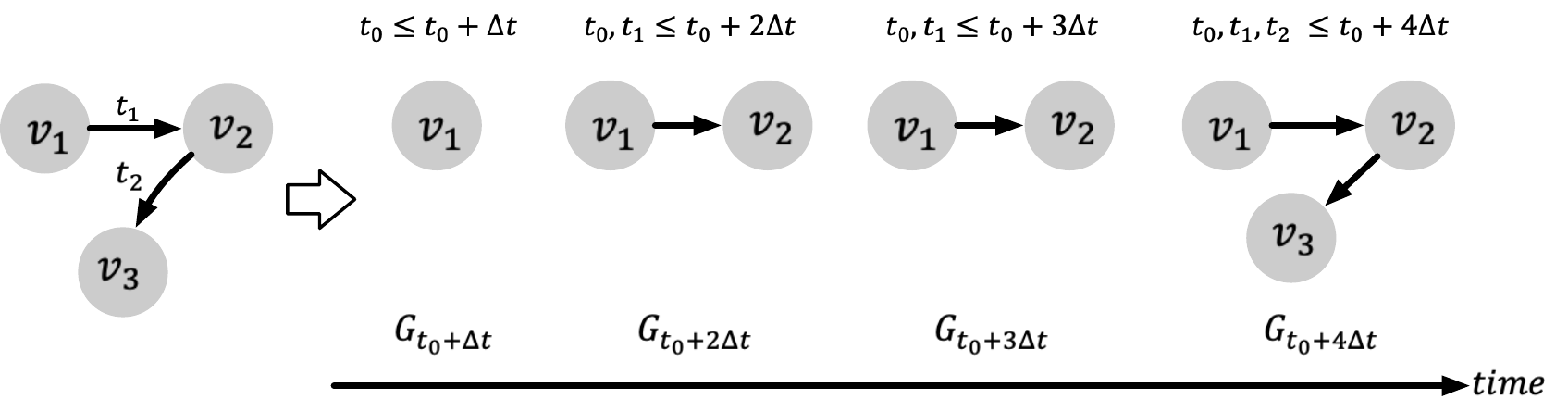}
    \caption{An illustration of t-Alignment.}
    \label{fig:t-Align}
\end{figure}

\subsubsection{\textbf{t-Alignment}}

As the step size of ODE solvers are not necessarily equal to the time interval of two updates of a temporal session graph,  ODE solvers can not be directly applied to the dynamic graph setting. To solve this problem, we propose \textit{t-Alignment}, a technique that updates the graph structure in time when solving the ODE function. An illustration of \textit{t-Alignment} is shown in Figure~\ref{fig:t-Align}. In specific, we assign each edge a timestamp $t$ to represent the time the edge appears. Given the initial time point $t_0$, the number of the integral time point $k$, and the step size of ODE solver $\Delta t$. At each integral time point, $t_0+k\times\Delta t$, we check all the edges and preserve the edges with timestamp $t$ where $t\le t_0+k\times\Delta t$, to form the current session graph, by

\begin{equation}
    \mathcal{G}_{t_0+k\times\Delta t}=\left(\mathcal{V}_{t_0:t_0+k\times\Delta t}, \mathcal{E}_{t_0:t_0+k\times\Delta t} \right )
\end{equation}

\noindent Where $\mathcal{G}_{t_0+k\times\Delta t}$ is the graph at timestamp $t_0+k\times\Delta t$, $\mathcal{V}_{t_0:t_0+k\times\Delta t}$ and $\mathcal{V}_{t_0:t_0+k\times\Delta t}$ are item set and edge set that exist between $t_0$ and $t_0+k\times\Delta t$ respectively. Then the GNG-ODE function simply takes $G_{t_0+k\times\Delta t}$ as input and infers $\frac{d\bm{h}(t_0+k\times\Delta t)}{dt}$, the hidden states of all items. In such a setting, we only need to solve the ODE function once to get the hidden state of the end of the session.  Besides, we do not need to store all snapshots of a temporal session graph or interrupt the integral process of ODE solvers to update the graph.


\subsection{User Preference Generation and Prediction}

After obtaining the item representations right after the last update time of a temporal session graph $G_{t_n}$,  which are denoted as $\bm{h}_{v_1}(t_n^+),... \bm{h}_{v_n}(t_n^+)$, we generate the hybrid preference representation to represent the current user interests. Specifically, we combine the recent interest and the long-term preference in the ongoing session to obtain the user’s preference. We use the vector of the last item as the recent interest, that is, $\hat{\bm{z}}_{r}=\bm{h}_{v_n}(t_n^+)$, where $\hat{\bm{z}}_{r}\in \mathbb{R}^{d}$.

For long-term interest, we consider all items in the session and utilize an attention mechanism to determine the weights for combining the historical item vectors, as follows:
\begin{equation}
    \begin{split}
    \hat{\bm{z}}_{l}&=\Sigma_{i=1}^{n}\bm{\gamma}_{i}\bm{h}_{v_i}(t_n^+) \\
    \bm{\gamma}_{i} &= \text{Softmax}(\bm{a}_{i}) \\
    \bm{a}_{i} &= W_{1}\sigma\left (W_{2}\bm{h}_{v_i}(t_n^+)+W_{3}\hat{\bm{z}}_{r}+\bm{b}\right )
    \label{eq:long_pref}
    \end{split}
\end{equation}

\noindent where $\hat{\bm{z}}_{l}\in \mathbb{R}^{d}$ is the generated long-term preference at the $t^*$-th timestamp, $\bm{a}_{i}$ and $\bm{\gamma}_{i}$ are the importance scores of item $\bm{v}_{i}$ before and after normalization, respectively, and $W_{1}\in \mathbb{R}^{d}, W_{2},W_{3}\in \mathbb{R}^{d\times d}$ are learnable parameters. $\sigma$ is the sigmoid function.

Then, we generate the dynamic hybrid user preference by taking into account the long-term and recent interests, which can be denoted as:
\vspace{-0.2cm}
\begin{equation}
    \begin{split}
    \hat{\bm{z}}_{h} =W_{4}\left [\hat{\bm{z}}_{l};\hat{\bm{z}}_{r} \right ],
    \label{eq:hybrid_pref}
    \end{split}
\end{equation}
where $\hat{\bm{z}}_{h}\in \mathbb{R}^{d}$ is the final generated user preference at the timestamp $t_n$, and $W_{4}\in \mathbb{R}^{d\times 2d}$ is the learnable parameters.

After that, we can make predictions by computing a probability distribution of the candidate items to be clicked at the next timestamp through the multiplication operation between the user preference and the embeddings of each item in $V$:
\begin{equation}
    \begin{split}
    \tilde{\bm{y}}_{i} =||\hat{\bm{z}}_{h}||^{T}||\bm{x}_{i}||,
    \label{eq:pred}
    \end{split}
\end{equation}
 \noindent where $||\cdot||$ denotes the $L_2$ normalization operation.
 
 Finally, we compute the normalized score for each candidate item, as follows: 
\vspace{-0.1cm}
\begin{equation}
    \hat{\bm{y}}_i=\text{Softmax}(\tilde{\bm{y}}_i)
    \label{eq:softmax}
\end{equation}

\noindent where $\hat{\textbf{y}}=\{\hat{\bm{y}}_{1},\hat{\bm{y}}_{2},...,\hat{\bm{y}}_{|V|}\}$ are the normalized prediction scores vector for all candidate items.

\subsection{Training}

After obtaining the preferential scores, we adopt cross-entropy as the optimization objective to learn the parameters following~\cite{chen2020handling,gupta2019niser,liu2018stamp,xia2021self}. The loss function is:
\vspace{-0.2cm}
\begin{equation}
  L(\hat{\textbf{y}})=-\sum_{i=1}^{|I|}\bm{y}_i\log(\hat{\bm{y}}_i)+(1-\bm{y}_i)\log(1-\hat{\bm{y}}_i) + \lambda ||\bm{\Theta}||_2^2
  \label{eq:15}
\end{equation}

\noindent where $\bm{y}_i\in \textbf{y}$ reflects the appearance of an item in the one-hot
encoding vector of the ground truth, \textit{i.e.,} $y_i=1$ if the $i$-th item is
the target item of the given session; otherwise, $\bm{y}_i=0$. $\bm{\Theta}$ is the model parameter. $\lambda$ is a scalar to control the influence of $L_2$ regularization. We use scaled softmax at the normalization stage Eq.~(\ref{eq:softmax}) to prevent over smoothing of relevance scores. 

\subsection{Computational Complexity Analysis}

Here we analyze the complexity of the initial latent state encoder and the dynamic representation learning module. Given the item set in the session as $V$, the transitions~(edges) in the session as $E$. Then the time complexity of initial latent state encoder GGNN is $O((|E|+|V|)l)$, where $l$ is the number of layers of GGNN. The time complexity of the dynamic representation learning module is $O((|E|+|V|)T/\Delta t)$ where $T$ is the time duration of the whole session and $\Delta t$ is the average integration step size of ODE solvers. Then the overall time complexity of the two modules is $O((l+T/\Delta t)(|E| + |V|))$, which is a linear combination of $|E|$ and $|V|$. As $l$ and $T/\Delta t$ are relatively small, we find that the total time complexity increases but is still acceptable.

 

\section{Experiment}
In this section, we have conducted extensive experiments, and analyzed the performance of the proposed GNG-ODE method by addressing the
following key research questions as follows:
\begin{itemize}
    \item \textbf{RQ1:} Can our proposed GNG-ODE outperform the state-of-the-art baselines for session-based recommendation?
    \item \textbf{RQ2:} How does GNG-ODE perform with different encoders?
    \item \textbf{RQ3:} How does GNG-ODE perform comparing with other Neural ODE model? Is \textit{t-Alignment} useful to help GNG-ODE jointly capture structural and temporal pattern?
    \item \textbf{RQ4:} How well does GNG-ODE perform with different ODE solvers from the effectiveness perspective?
    \item \textbf{RQ5:} How is the scalability of GNG-ODE?
    \item \textbf{RQ6:} How do different hyper-parameters affect GNG-ODE?
\end{itemize}

\begin{table}[h]
    \centering
    \caption{Statistics of datasets.}
    \resizebox{.8\linewidth}{!}{
    \begin{tabular}{cccc}
        \toprule
                & Gowalla & Tmall & Nowplaying \\
        \midrule
        \#clicks & 1,122,788 & 818,479 & 1,367,963  \\
        \#train sessions   & 675,561 & 351,268 & 825,304  \\
        \#test sessions  & 155,332 & 25,898 & 89,824  \\
        \#items  & 29,510 &  40,727 & 60,416  \\
        Average length  & 4.32 & 6.69  & 7.42  \\
        Average Interval & 11.07h & 1.49s & 4.36h \\
        \bottomrule
    \end{tabular}}
    \label{tab:dataset}
\end{table}

\begin{table*}[htbp!]
\Large
\centering
  \caption{Results(\%) of main experiments. The numbers of HR@10, HR@20, MRR@10 and MRR@20 are reported. $\ast$ denotes a significant improvement of GNG-ODE over the best baseline using a paired $t$-test ($p$ < 0.01).}
  \label{tab:main}
  \resizebox{.8\linewidth}{!}{
  \begin{tabular}{c|cccc|cccc|cccc}
    \toprule
    \multirow{2}{*}{Model} & \multicolumn{4}{c}{Gowalla} & \multicolumn{4}{c}{Tmall} & \multicolumn{4}{c}{Nowplaying} \\
               ~&HR@10 & HR@20 & MRR@10 &MRR@20 & HR@10 & HR@20 & MRR@10 & MRR@20 & HR@10 & HR@20 & MRR@10 &  MRR@20 \\
    \midrule
    NARM   & $40.71$ & $49.84$ & $22.19$ & $23.63$ & $21.97$ & $27.14$ & $10.50$ & $12.89$ &  $13.11$ & $17.36$ & $6.03$ & $6.49$ \\
    SR-GNN & $42.18$ & $50.21$ & $23.63$ & $24.08$ & $24.79$ & $29.39$ & $13.61$ & $13.93$ &  $13.73$ & $18.46$ & $6.92$ & $7.25$ \\
    NISER+ & $44.92$ & $53.58$ & $25.03$ & $25.43$ & $29.81$ & $36.32$ & $15.27$ & $15.72$ &  $16.37$ & $22.35$ &  $8.05$ & $8.52$ \\
    SGNN-HN & $42.19$ & $50.92$ & $24.10$ & $25.26$ & $22.09$ & $29.20$ & $11.85$ & $15.40$ & $13.22$ & $17.52$ & $7.15$ & $7.41$ \\
    LESSR & $42.87$ & $51.04$ & $23.90$ & $24.47$ & $27.93$ & $32.75$ & $14.79$ & $14.91$  & $14.73$ & $19.54$ & $7.66$ & $7.99$  \\
    GCE-GNN & $45.38$ & $53.77$ & $25.11$ & $\underline{25.69}$ & $28.01$ & $33.42$ & $15.08$ & $15.42$ & $16.94$ & $22.37$ & $8.03$ & $8.40$ \\
    DAT-MDI & $\underline{45.46}$ & $53.98$ & $\underline{25.14}$ & $25.65$ & $27.98$ & $33.23$ & $15.21$ & $15.64$ & $\underline{16.98}$ & $\underline{22.45}$ & $8.12$ & $8.67$ \\
    \midrule
    TiSASRec & $44.67$ & $53.39$ & $24.06$ &  $24.46$  & $28.23$ & $35.28$ & $15.67$ & 16.02 & $16.56$ & $21.28$ & $7.14$ & $7.69$ \\
    TGSRec & $44.98$ & $53.89$ & $24.89$ & $25.12$  & $29.54$ & $36.43$ & $15.99$ & $16.37$ & $16.67$ & $21.56$ & $7.34$ & $7.87$ \\
    TMI-GNN & $45.43$ & $\underline{54.03}$ & $25.12$ & $25.47$ & $\underline{30.06}$ & $\underline{36.87}$ & $\underline{16.94}$ & $\underline{17.19}$ & $16.89$ & $22.28$ & $\underline{8.39}$ & $\underline{8.89}$ \\
    \midrule
    GNG-ODE & $\textbf{46.05}^{\ast}$ & $\textbf{54.58}^{\ast}$ & $\textbf{26.32}^{\ast}$ & $\textbf{26.91}^{\ast}$ & $\textbf{31.11}^{\ast}$ & $\textbf{37.66}^{\ast}$ & $\textbf{17.90}^{\ast}$ &  $\textbf{18.23}^{\ast}$ & $\textbf{17.31}^{\ast}$ &  $\textbf{22.83}^{\ast}$ & $\textbf{8.85}^{\ast}$ & $\textbf{9.23}^{\ast}$ \\
    \midrule 
    \textit{Improv.} & $1.30\%$ & $1.02\%$ & $4.69\%$ & $4.75\%$ & $3.49\%$ & $2.14\%$ & $5.67\%$ & $6.05\%$ & $1.94\%$ & $1.69\%$ & $5.48\%$ & $3.82\%$ \\
    \bottomrule
  \end{tabular}
  }
\end{table*}
\vspace{-0.1cm}

\subsection{Datasets and Preprocessing}
We evaluate GNG-ODE and the baselines on the following three publicly available benchmark datasets, which are commonly used in the literature of session-based recommendation~\cite{Li2017NeuralAS,Qiu2019RethinkingTI,ren2019repeatnet,wu2019session,yuan2019simple,chen2020handling,Xu2019GraphCS,Pan2020StarGN,gupta2019niser}:
\begin{itemize}
    \item \textit{Gowalla}\footnote{\noindent  \url{https://snap.stanford.edu/data/loc-gowalla.html}} is a dataset that contains users' check-in information for point-of-interest recommendation. Following ~\cite{guo2019streaming,tang2018personalized,chen2020handling}, we keep the 30,000 most popular locations and set the splitting interval to 1 day, and consider the last {20\%} of sessions for testing.
    \item \textit{Tmall}\footnote{\noindent   \url{http://ocelma.net/MusicRecommendationDataset/lastfm-1K.html}} is a user-purchase data (only purchase records are utilized) obtained from Tmall platform. We also use the last {20\%} of sessions as the test sets.
    \item \textit{Nowplaying}\footnote{\noindent   \url{https://dbis.uibk.ac.at/node/263\#nowplaying}~\cite{poddar2018nowplaying}} is a comprehensive implicit feedback dataset consisting of user-song interactions crawled from Twitter. For each music, we randomly select 20\% of users who have played the music as the test sets, and the remaining users for training.
\end{itemize}

Following~\cite{Li2017NeuralAS,liu2018stamp,Qiu2019RethinkingTI,ren2019repeatnet,wu2019session,chen2020handling,zhang2022efficiently}, we filter the sessions containing merely an item and the items appearing less than five times for each dataset. We further make data augmentation that has been widely applied in ~\cite{Li2017NeuralAS,liu2018stamp,wu2019session,chen2020handling} after filtering short sessions and infrequent items. The statistics of the datasets are shown in Table~\ref{tab:dataset}.


\subsection{Baseline Models} We consider the following baselines to evaluate the performance of the proposed model.
\begin{itemize}
    \item \textbf{NARM}\footnote{\noindent \url{https://github.com/lijingsdu/sessionRec\_NARM}} ~\cite{Li2017NeuralAS} is a RNN-based method for session-based recommendation. It utilizes RNNs with attention to model user action sequences.
    \item \textbf{SR-GNN}\footnote{\noindent \url{https://github.com/CRIPAC-DIG/SR-GNN}} ~\cite{wu2019session} is a GNN-based method for session-based recommendation. It applies GNNs to extract item features and obtains session representaion through traditional attention network ~\cite{li2015gated}.
    \item \textbf{NISER+}\footnote{\noindent \url{https://github.com/johnny12150/NISER}} ~\cite{gupta2019niser} employs normalized item and session embedding based on graph neural network to alleviate popularity bias problem in session-based recommendation.
    \item \textbf{SGNN-HN} ~\cite{Pan2020StarGN} solves the long-range information propagation problem by adding a star node to take the non-adjacent items into consideration with gated graph neural networks.
    \item \textbf{LESSR}\footnote{\noindent \url{https://github.com/twchen/lessr}} ~\cite{chen2020handling} transforms the sessions into directed multigraphs and propagates information along shortcut connections to solve the lossy session encoding problem.
    \item \textbf{GCE-GNN}\footnote{\noindent \url{https://github.com/CCIIPLab/GCE-GNN}} ~\cite{wang2020global} transforms the sessions into global graph and local graphs to enable cross session learning.
    \item \textbf{DAT-MDI}~\cite{chen2021dual} combines the GNN and GRU to learn the cross session enhanced session representation.
    \item \textbf{TiSASRec}~\cite{li2020time} introduces to use relationship matrix to model the temporal relations for items in the sequence. 
    \item \textbf{TGSRec}~\cite{fan2021continuous} uses a user collaboratively continuous-time transformer for sequential recommendation.  
    \item \textbf{TMI-GNN}~\cite{shen2021temporal} uses temporal information to guide the multi-interest network to capture more accurate user interests.
\end{itemize}


\subsection{Implementation}
We apply grid search to find the optimal hyper-parameters for each model.
We use the last 20\% of the training set as the validation set. The ranges of other hyper-parameters are $\{64, 128, 256,512\}$ for hidden dimensionality $d$ and $1e{-3}$ chosen for learning rate $\eta$. The weight decay rate $\lambda$ is set to $1e{-4}$. We use the Adam optimizer to train the models. The batch size  is set to 512. We run all models five times with different random seeds and report the average.  We use the same evaluation metrics \textbf{HR@K} (Hit Rate) and \textbf{MRR@K} (Mean Reciprocal Rank) following previous studies ~\cite{Li2017NeuralAS,Qiu2019RethinkingTI,ren2019repeatnet,wu2019session,chen2020handling,Xu2019GraphCS,Pan2020StarGN,gupta2019niser}. The implementation of our model can be found at~\url{https://github.com/SpaceLearner/GNG-ODE}.


\subsection{Overall Comparison (RQ1)}

To demonstrate the overall performance of the proposed model, we compare it with the state-of-the-art recommendation methods. They include the static models NARM, SR-GNN, NISER+, SGNN-HN, LESSR, GCE-GNN and DAT-MDI, and temporal models like TiSASRec, TGSRec and TMI-GNN.  The experimental results of all compared methods are shown in Table~\ref{tab:main}, from which we have the following observations.


Compared with RNN, GNN-based models have a stronger ability to explore complex graph-structured data. Moreover, LESSR works better than SR-GNN and SGNN-HN, demonstrating that handling the lossy session encoding problem can further help boost the recommendation performance of GNN models. In addition, through exploiting the global-level transitions between items, GCE-GNN outperforms many baselines on Gowalla that achieves the best performance in the static baselines. However, the performance of GCE-GNN on two other datasets is not satisfactory compared with NISER+, indicating that the long-tail problem and the overfitting problem are more prevalent outside the check-in scenario dataset (Gowalla). DAT-MDI performs the best on Gowalla dataset, which verifies the importance of capturing the complex structural pattern across sessions. We also observe that temporal information helps capture user preference, as temporal baselines all achieve comparable performance. Among temporal baselines, TMI-GNN performs the best, indicating that decomposing temporal information into different interests captures more fine-grained user preference.


Next, we zoom in on the performance of our proposed GNG-ODE. First, we can observe that GNG-ODE can achieve state-of-the-art performance for all cases on three datasets. In particular, GNG-ODE outperforms the existing temporal baselines (\textit{i.e.,} TiSASRec, TGSRec and TMI-GNN). We attribute the improvements of GNG-ODE against the baselines to two factors: One is that GNG-ODE can take the continuous evolution of the session graph structures into consideration, and the other one is that GNG-ODE solves the continuous concept modeling problem using the continuous ODE function. In addition, the improvements of GNG-ODE over the best baselines (\textit{i.e.,} DAT-MDI, GCE-GNN and TMI-GNN) in terms of HR@20 and MRR@20 are 2.14\% and 6.05\% on Tmall, respectively, and the corresponding improvements are 1.69\% and 3.82\% on the Nowplaying dataset. 
We can observe that on all  datasets, 
GNG-ODE brings more performance gain over original models when K in evaluation when HR@K is smaller. A small value of K means the target items stay in the top  positions of the recommendation list. Due to the position bias ~\cite{chen2020bias} in recommendation that users tend to pay more attention to the items in a higher position of the recommendation list, our framework can help original models to produce more accurate and user-friendly recommendations.

\subsection{Impact of Encoders for Initial State Inference (RQ2)}
In this experiment, we compare GNG-ODE with different initial state encoders to investigate the contribution of our encoder design. The following variants are tested on all datasets, where the results are reported in Table~\ref{tab:encoders}:
\begin{enumerate}
    \item Identity: GNG-ODE with raw one-hot embeddings as the initial hidden state.
    \item MLP: GNG-ODE with the output of a two-layer MLP as the initial hidden state.
    \item GGNN: GNG-ODE with the output of a GGNN as the initial hidden state.
\end{enumerate}

\begin{table}[tbp]
    \centering
    \caption{Results of Different Initial State Encoders.}
    \label{tab:encoders}
    \resizebox{\linewidth}{!}{
    \Large
    \begin{tabular}{c|cc|cc|cc}
     \toprule
      Dataset  & \multicolumn{2}{c}{Gowalla} & \multicolumn{2}{c}{Tmall} & \multicolumn{2}{c}{Nowplaying} \\
      Metrics & HR@20 & MRR@20 & HR@20 & MRR@20 & HR@20 & MRR@20 \\
      \midrule
      Identity & 53.84 & 26.80 & 35.03 & 15.11 & 22.75 & 9.23 \\
      MLP & 53.86 & 26.35 & 34.75 & 14.58 & 21.45 & 8.39 \\
      GGNN & \textbf{54.58} & \textbf{26.91} & \textbf{37.66} & \textbf{17.25} & \textbf{22.83} & \textbf{9.45} \\
      
      \bottomrule
    \end{tabular}
    }
\end{table}
From Table~\ref{tab:encoders}, we can observe that GGNN achieves best perforamnce on all the three datasets. GGNN encoder emphasizes more on capturing structural information, replacing this with raw embedding or MLP will significantly decrease the recommendation performance on the Tmall dataset. For Gowalla and Nowplaying, compared with the results on Tmall,  the structural information contributes less on both HR@20 and MRR@20 metrics. Our analysis is that the difference may be caused by how the influence of the structural and temporal factors in the e-commerce and check-in as well as interest-based scenarios varies. Specifically, in the e-commerce dataset, \textit{i.e.,} Tmall, the structural information is relatively more important, since the transition relation between items is much more complicated than the simple sequential signal~\cite{wu2019session,Qiu2019RethinkingTI}. 


\subsection{Impact of ODE Functions (RQ3)}
To verify the effectiveness of GNG-ODE and \textit{t-Alignment}, we replace the GNG-ODE with several widely used ODE functions and compare their recommendation performance. The variants of GNG-ODE are listed as follows. For none t-Alignment version, we use the static session graph as input.
\begin{enumerate}
    \item GNG-ODE: the model proposed in this paper. 
    \item GCN-ODE: use a two layer graph convolutional network~\cite{DBLP:conf/iclr/KipfW17} as the ODE function.
    \item GRU-ODE: use a one layer gated recurrent unit~\cite{cho2014learning} as the ODE function.
    \item MLP-ODE: use a two layer linear network with GELU acivation~\cite{Hendrycks2016GaussianEL} as the ODE function.
\end{enumerate}

Table~\ref{tab:t-Align} shows the performance of different GNG-ODE variants. We observe that without t-Alignment the performance will degrade on all datasets. This confirms that building continuous session graphs could enable our model to capture the evolution of the session graphs over multi-time steps. This further demonstrates the indispensability of t-Alignment in expanding the applicability of those ODE solvers. Moreover, we find that the method jointly considering structural and temporal patterns~(GNG-ODE) outperforms its counterparts that only consider structural patterns~(GCN-ODE) or that only consider temporal patterns~(GRU-ODE), demonstrating the superiority of GNG-ODE at capturing both information.
 
 \begin{table}[tbp]
    \centering
    \caption{Impact of ODE Function Module.}
    \label{tab:t-Align}
    \resizebox{\linewidth}{!}{
    \Large
    \begin{tabular}{c|cc|cc|cc}
     \toprule
      Dataset  & \multicolumn{2}{c}{Gowalla} & \multicolumn{2}{c}{Tmall} & \multicolumn{2}{c}{Nowplaying} \\
      Metrics & HR@20 & MRR@20 & HR@20 & MRR@20 & HR@20 & MRR@20 \\
      \midrule
      GNG-ODE & \textbf{54.58} & \textbf{26.91} & \textbf{37.66} & \textbf{18.23} & \textbf{22.83} & \textbf{9.45} \\
      GCN-ODE & 54.46 & 26.63 & 37.45 & 17.83 & 22.65 & 9.25 \\
      GRU-ODE & 54.41 & 26.76 & 37.33 & 17.30 & 22.59 & 9.23 \\
      MLP-ODE & 54.34 & 26.81 & 37.17 & 17.63 & 22.34 & 9.18 \\
      \bottomrule
      GNG-ODE w/o t-Align & 54.16 & 25.76 & 37.22 & 17.86 & 22.70 & 9.17 \\
      GCN-ODE w/o t-Align & 53.85 & 25.57 & 37.18 & 17.37 & 22.62 & 9.14 \\
      GRU-ODE w/o t-Align & 54.19 & 26.25 & 36.88 & 17.28 & 22.36 & 9.06 \\
      MLP-ODE w/o t-Align & 53.99 & 26.12 & 36.89 & 17.36 & 22.10 & 9.04 \\
      \bottomrule
    \end{tabular}}
    \vspace{-0.1cm}
\end{table}

\subsection{Analysis of ODE Solvers (RQ4)}

In this section, we investigate the effect of different ODE solvers. ODE solvers play a central role in the performance of GNG-ODE. Widely used numerical ODE solvers are fixed-step solvers like explicit Euler~(Euler), fourth-order Runge–Kutta~(RK4) and adaptive step size solvers like Dopri5.


\begin{figure}[tbp]
\subfigure[Gowalla]{\centering
    \includegraphics[width=0.31\linewidth]{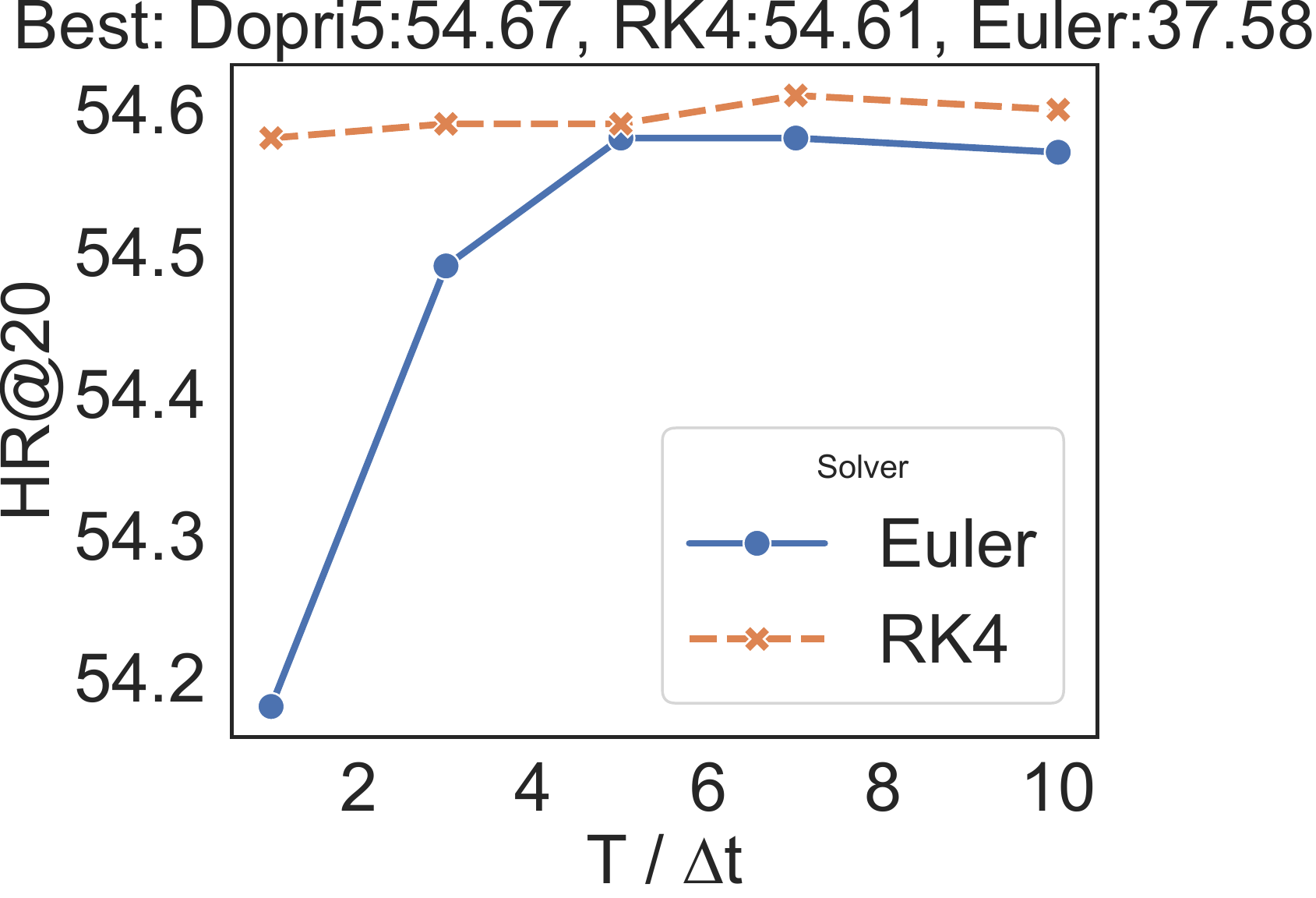}
    \label{fig:gowalla}}
\subfigure[Tmall]{\centering
    \includegraphics[width=0.31\linewidth]{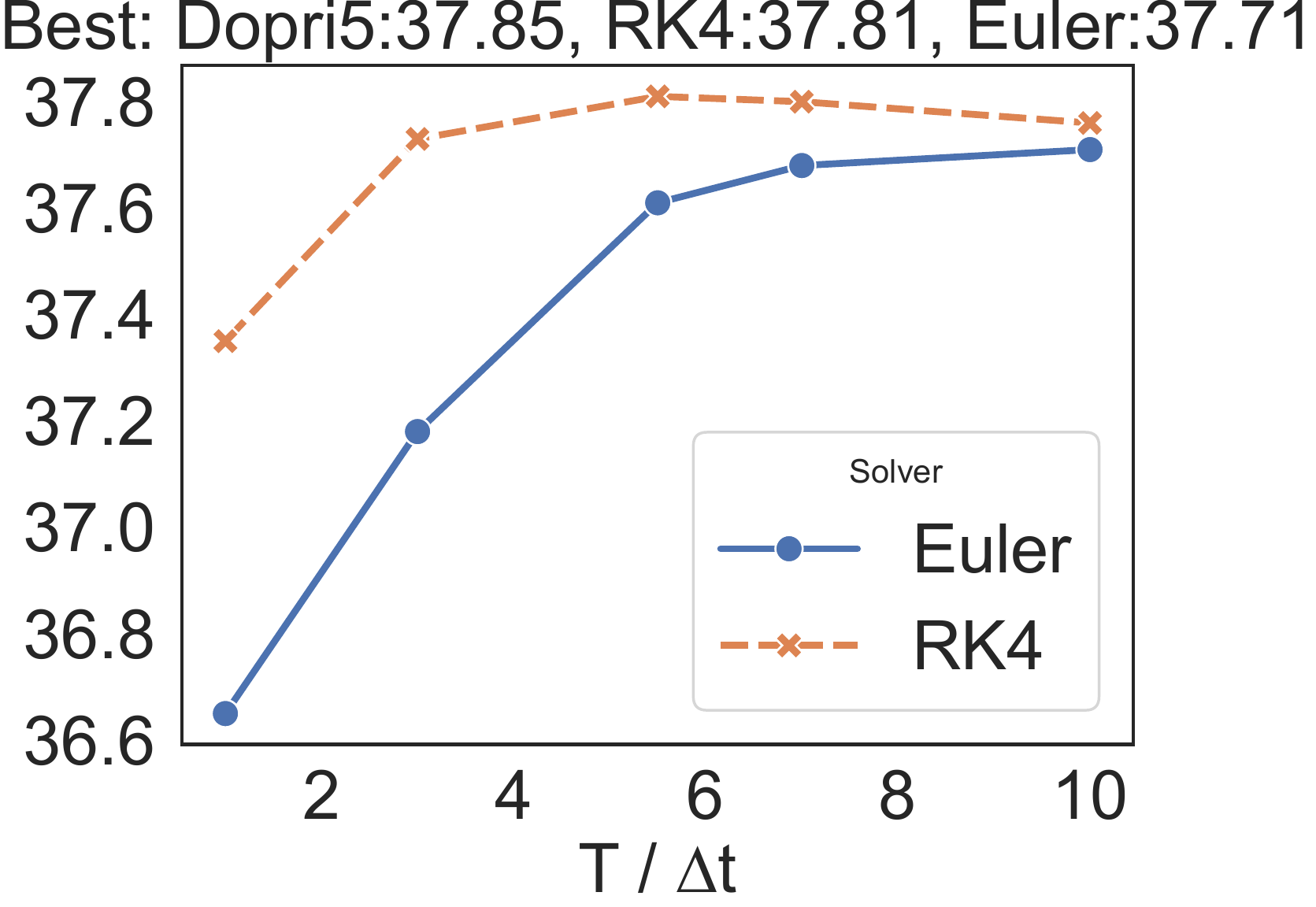}
    \label{fig:tmall}}
\subfigure[Nowplaying]{\centering
    \includegraphics[width=0.31\linewidth]{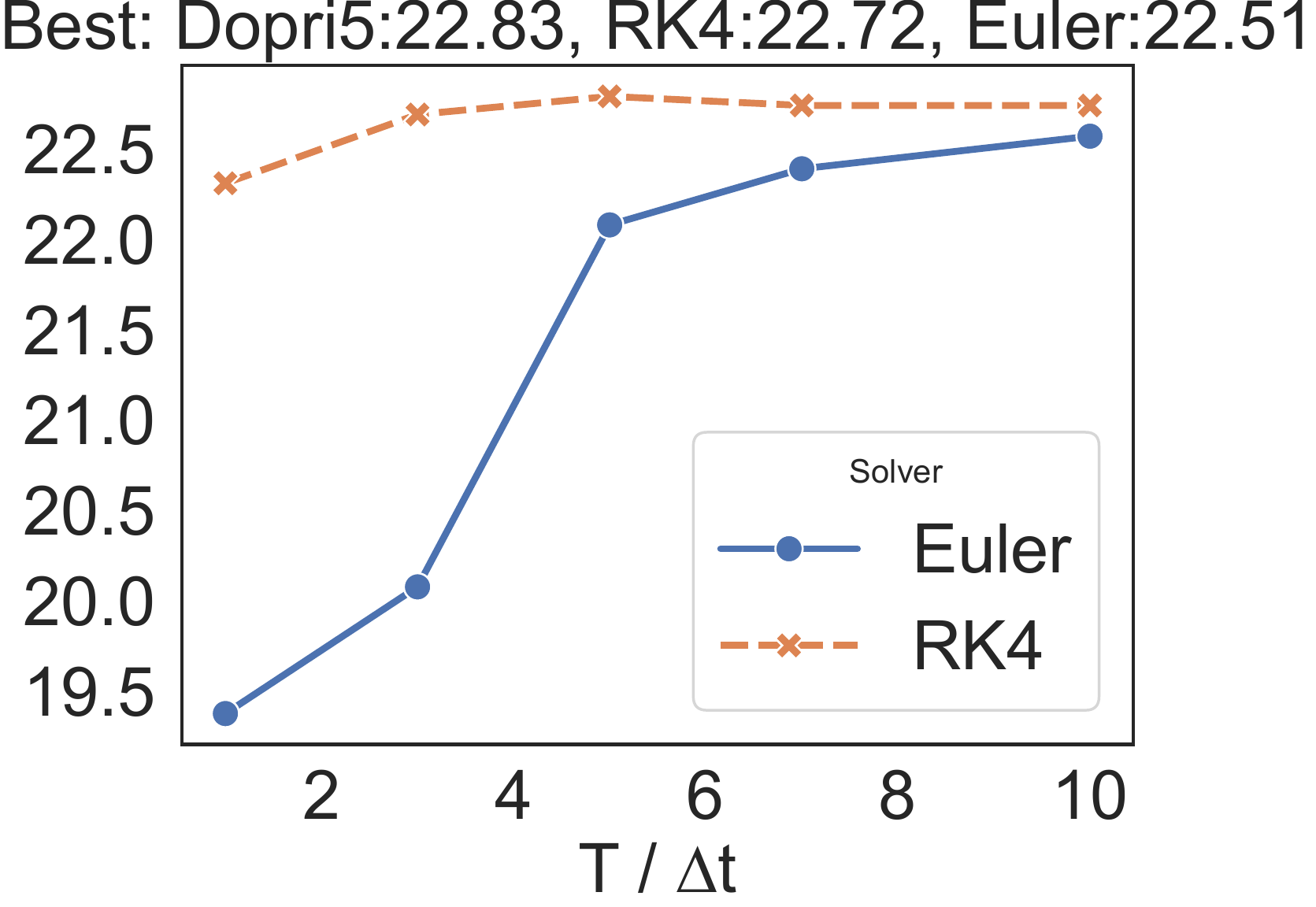}
    \label{fig:nowplaying}}
\subfigure[Gowalla]{\centering
    \includegraphics[width=0.31\linewidth]{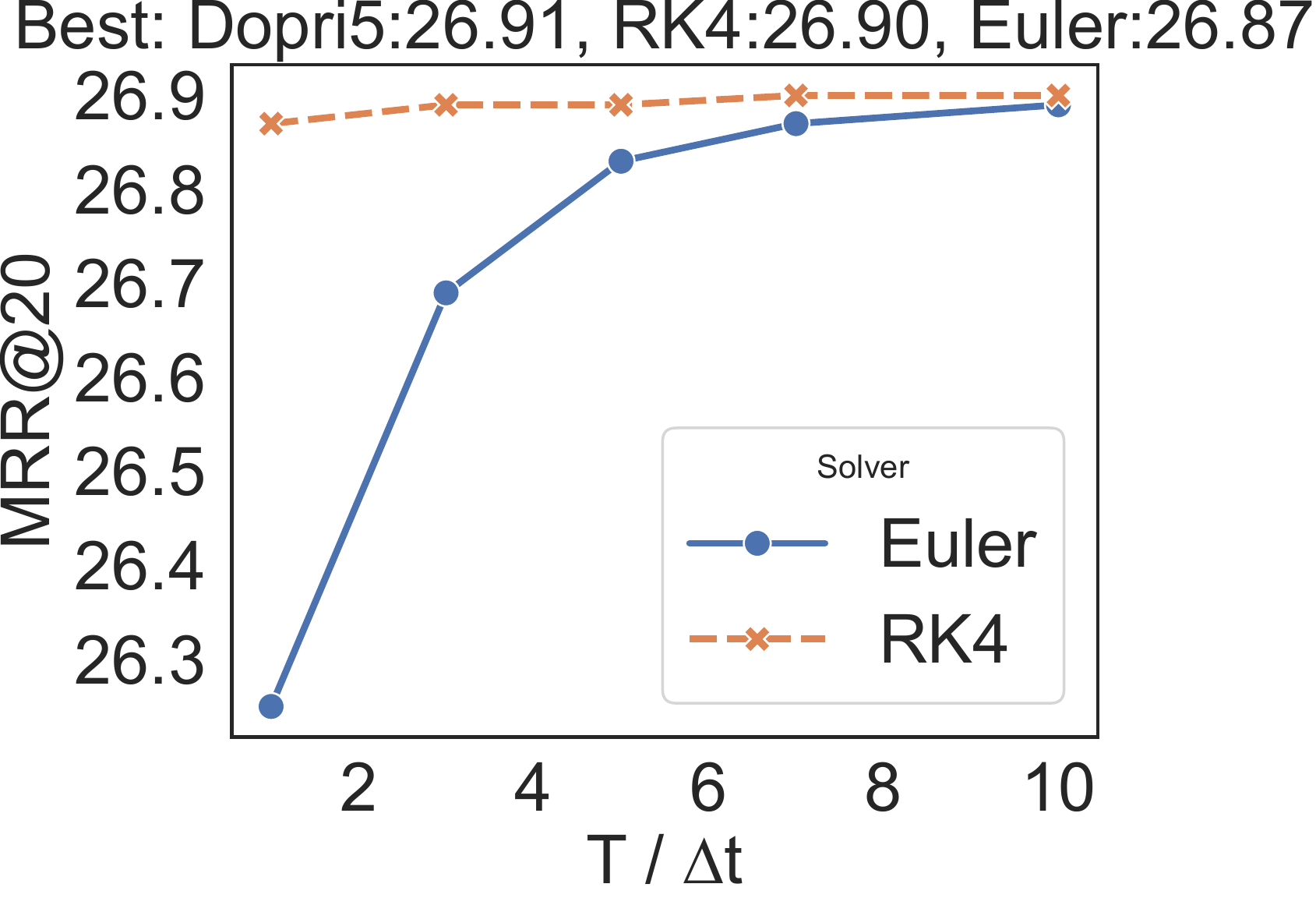}
    \label{fig:gowalla}}
\subfigure[Tmall]{\centering
    \includegraphics[width=0.31\linewidth]{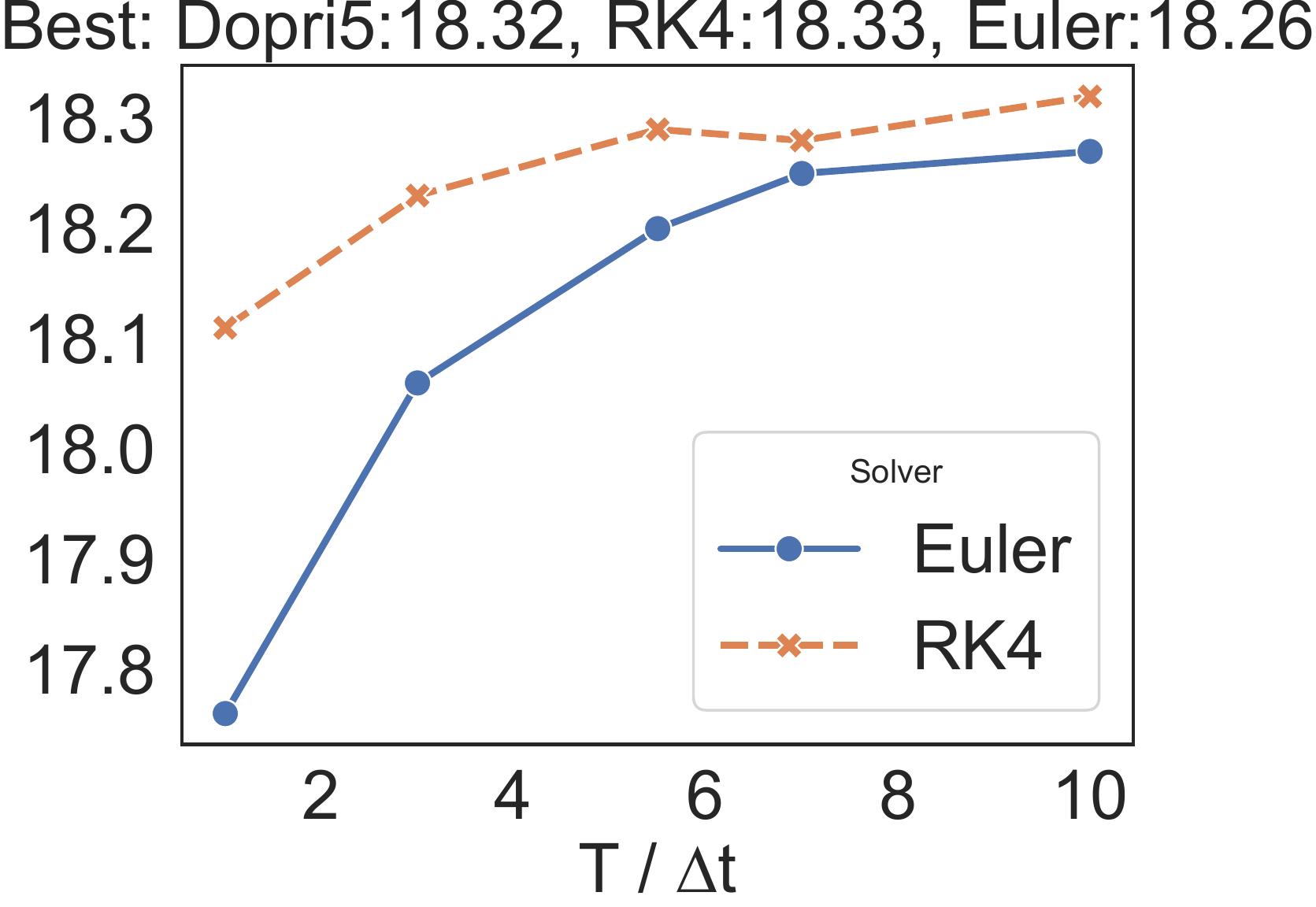}
    \label{fig:tmall}}
\subfigure[Nowplaying]{\centering
    \includegraphics[width=0.31\linewidth]{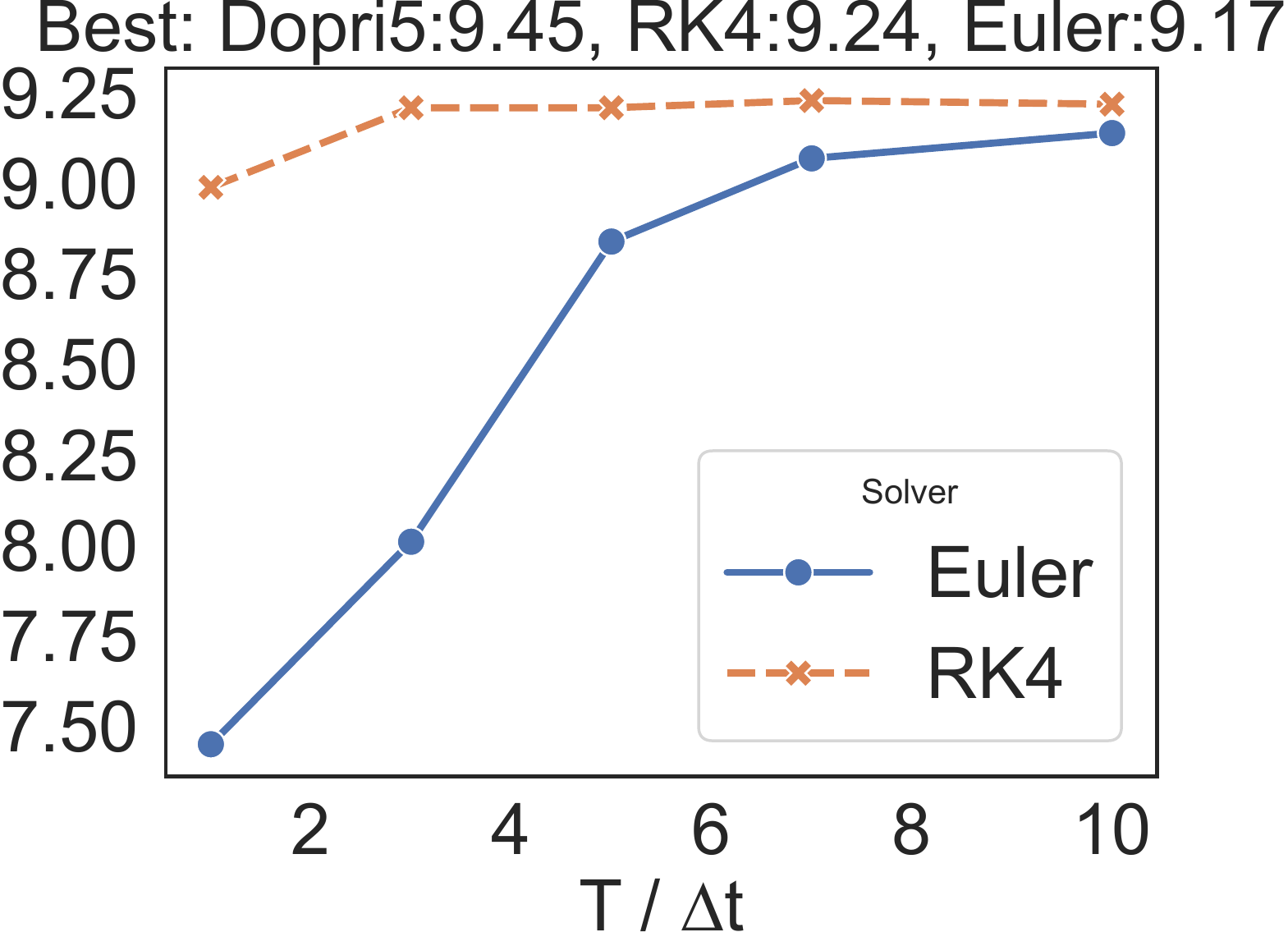}
    \label{fig:nowplaying}}
    \caption{The impact of ODE solvers.}
    \label{fig:solvers}
    \vspace{-0.1cm}
\end{figure}

\subsubsection{\textbf{Influence of ODE Solvers}}

The numbers of the best results for each solver are listed on the head of each subfigure in Figure~\ref{fig:solvers}. Adaptive step solver Dopri5 outperforms fixed-step solvers on the three datasets as adaptive step solvers adjust integration steps more flexibly than fixed step solvers. Besides, as RK4 takes more evaluation times in one step to achieve a more accurate estimation of the ODE function, it outperforms Euler on the three datasets.

\subsubsection{\textbf{Influence of Integration Step Size of Fixed Step ODE Solvers}}

Figure~\ref{fig:solvers} also summarizes the curves of fixed-step ODE solvers induced by various step sizes. The $x$-axis indicates the multiple of the time duration of a session relative to the integration step size. It is used as a hyper-parameter of the fixed step ODE solvers, \textit{e.g.,} Euler and RK4. We find that as the proportion of step size goes smaller, the performance of Euler and RK4 increases. It is because the estimated ODE function is more accurate in small step sizes. Besides, Euler is more sensitive than RK4 to step size as RK4 takes more estimation times in one integration step to ensure accuracy. Moreover, as the metric curves tend to converge as step size goes smaller, we do not need a too small step size to achieve good performance. To avoid taking too much running time, we can adjust the step size to balance both effectiveness and efficiency.


\subsection{Running Time Comparison~(RQ5)}

\begin{figure}[t]
    \centering
    \includegraphics[width=.9\linewidth]{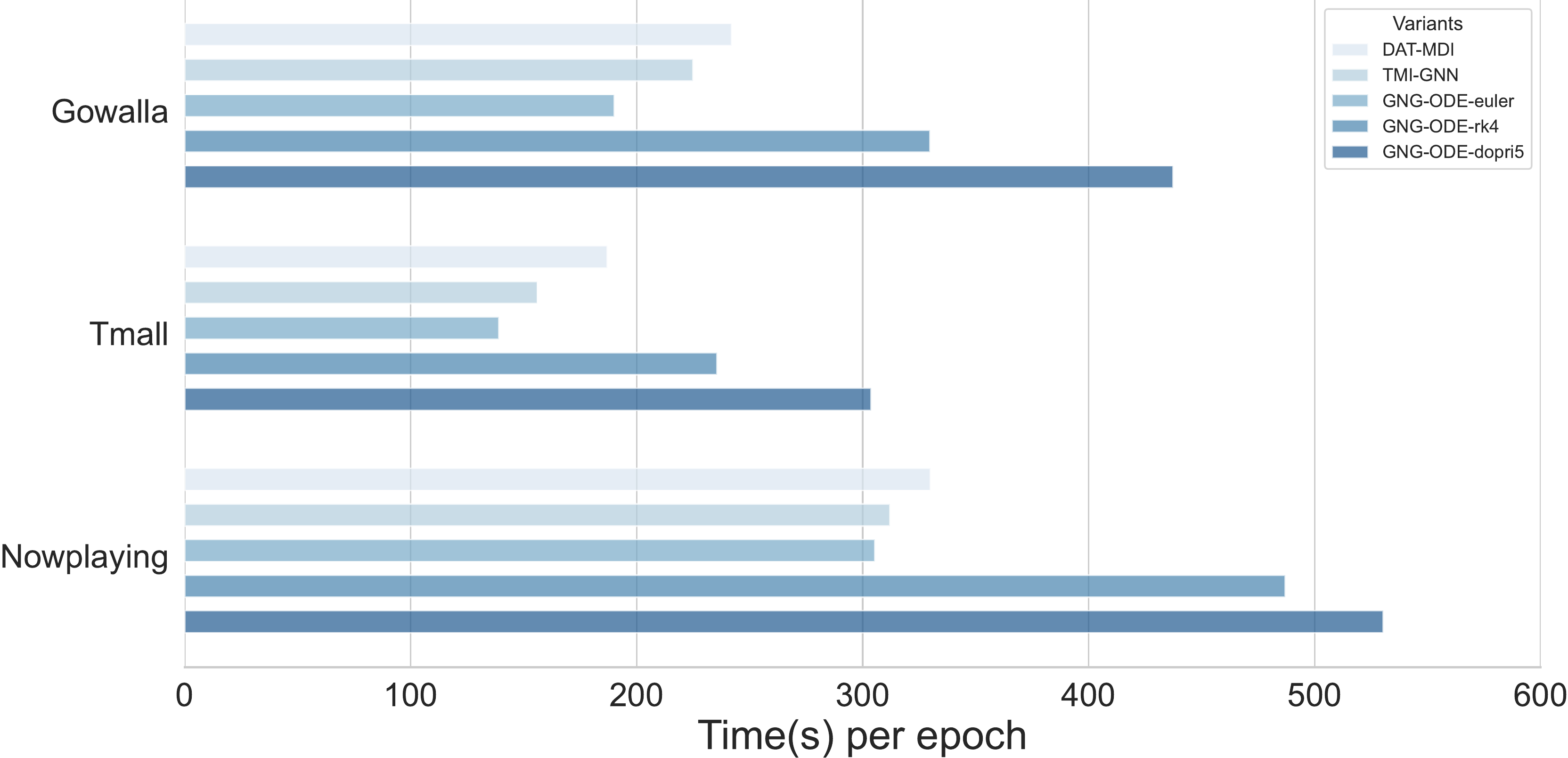}
    \caption{Running Time Comparison.}
    \label{fig:time}
    \vspace{-0.2cm}
\end{figure}

The computation time per epoch for GNG-ODE is summarized in Figure~\ref{fig:time}. We also give the running time of DAT-MDI, one recent baseline that does not consider temporal information and TMI-GNN, the best baseline that considers temporal information. For fix-step solvers we choose $T/\Delta t=7$. We find that the efficiency of GNG-ODE with Euler solver is on par with DAT-MDI and TMI-GNN. Although RK4 and Dopri5 take more time to compute, they achieve better performance and the time costs are still acceptable. 

\subsection{Hyper-parameter Study~(RQ6)}
To answer RQ6, we conduct experiments to study the sensitivity of GNG-ODE on the embedding dimension and the GGNN encoder layers. Specifically, we tune the the embedding dimension in \{64, 128, 256, 512\} and search the GGNN encoder layers in \{1,2,3,4,5\}. The ODE solver is set to RK4. The performance of GNG-ODE with different hyper-parameters is presented in Figure \ref{fig:hyper}.
\subsubsection{\textbf{Embedding Size}}
From Figure~\ref{fig:hyper}, we can observe that when the number of encoder layers is small, increasing the embedding dimension can generally improve the recommendation performance, especially from dimensions 64 to 128. This is because a large embedding dimension has a relatively better representation ability of item characteristics. However, there is a merely limited promotion of the performance when the dimension increases from 256 to 512. As increasing the embedding dimension will consume more computation resources, the dimension 256 is a proper choice considering both the effectiveness and efficiency of the recommender.

\subsubsection{\textbf{Number of GGNN Encoder Layers}}
\begin{figure}[tbp]
\subfigure[Gowalla]{\centering
    \includegraphics[width=0.31\linewidth]{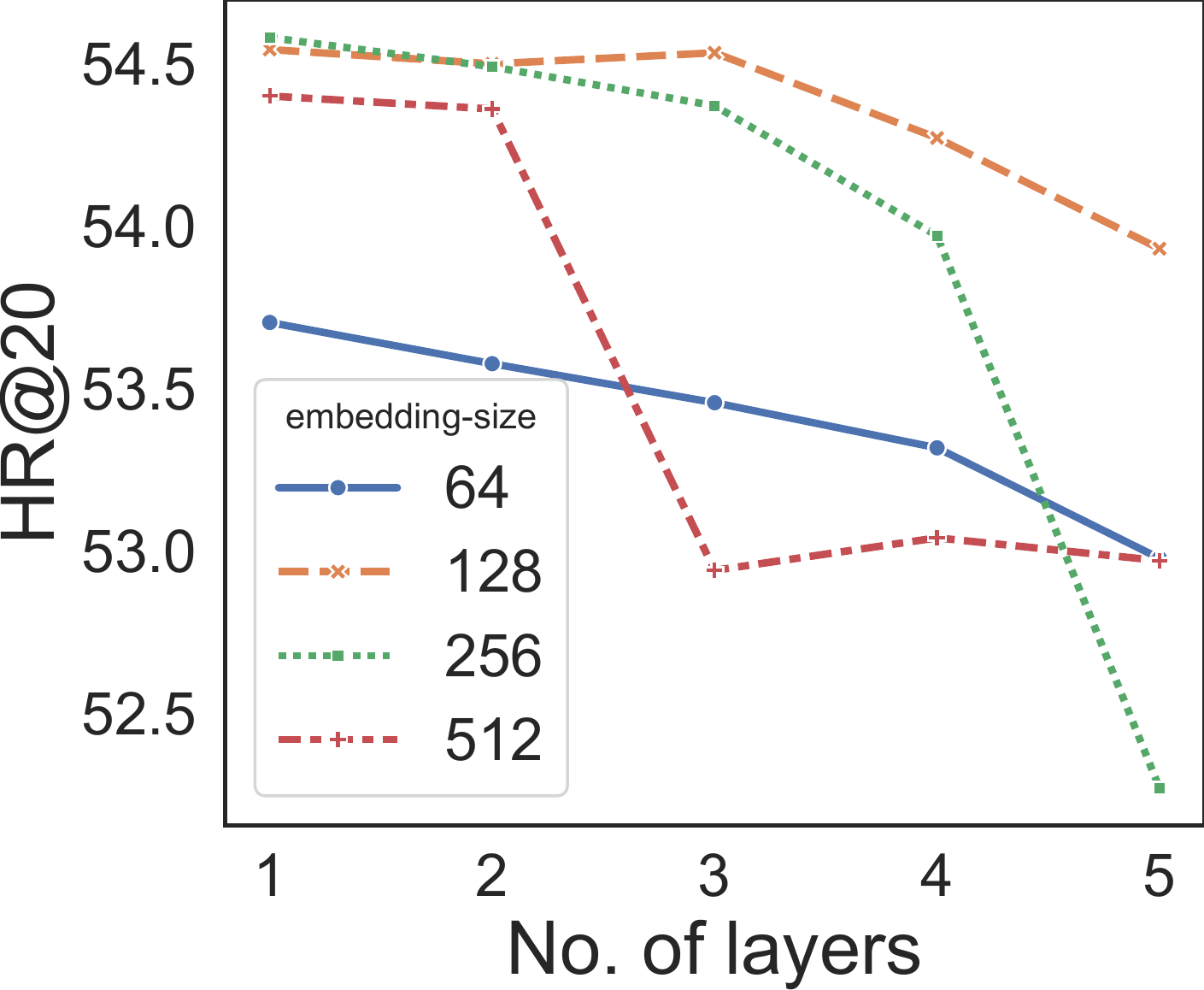}
    \label{fig:gowalla}}
\subfigure[Tmall]{\centering
    \includegraphics[width=0.31\linewidth]{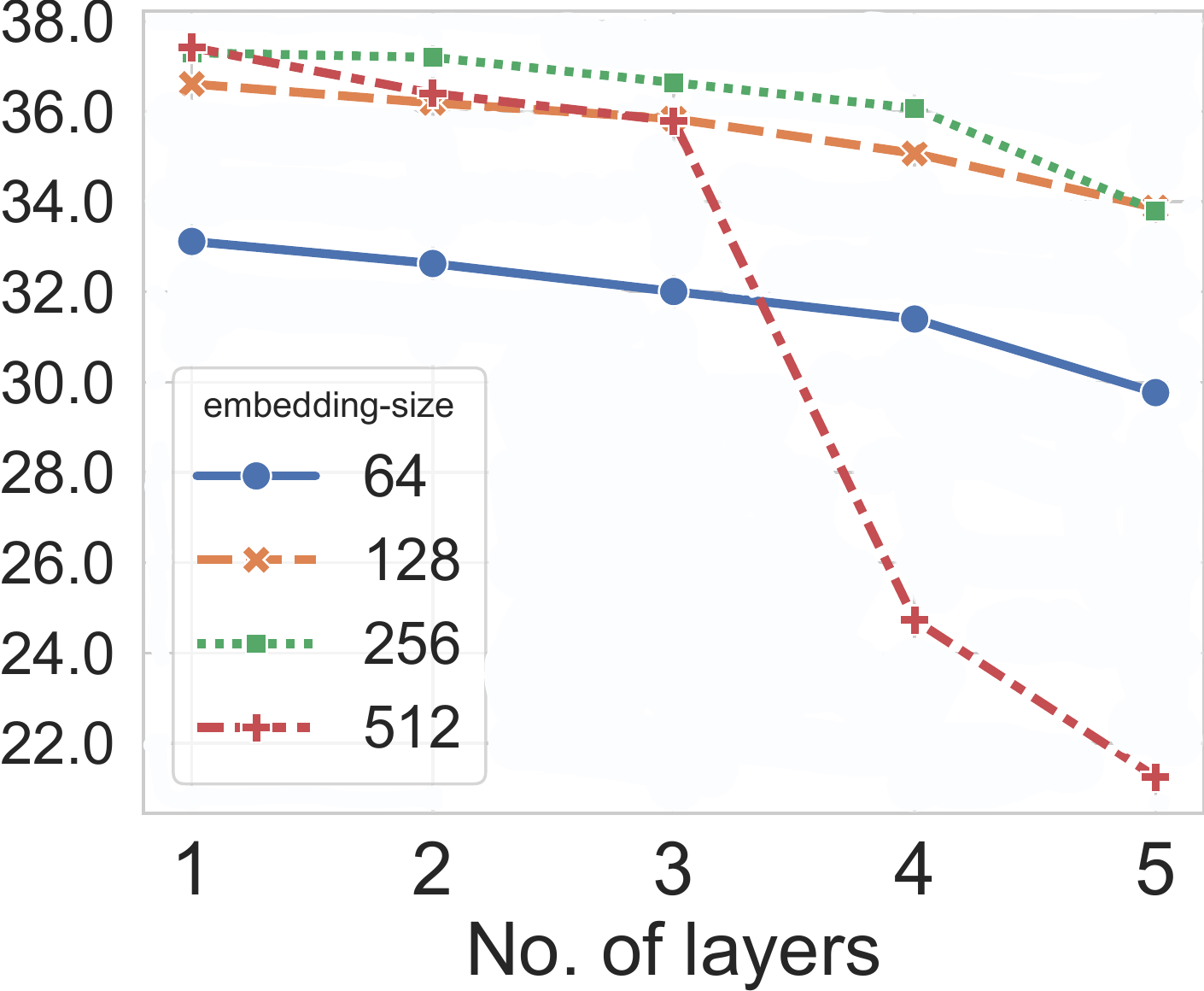}
    \label{fig:tmall}}
\subfigure[Nowplaying]{\centering
    \includegraphics[width=0.31\linewidth]{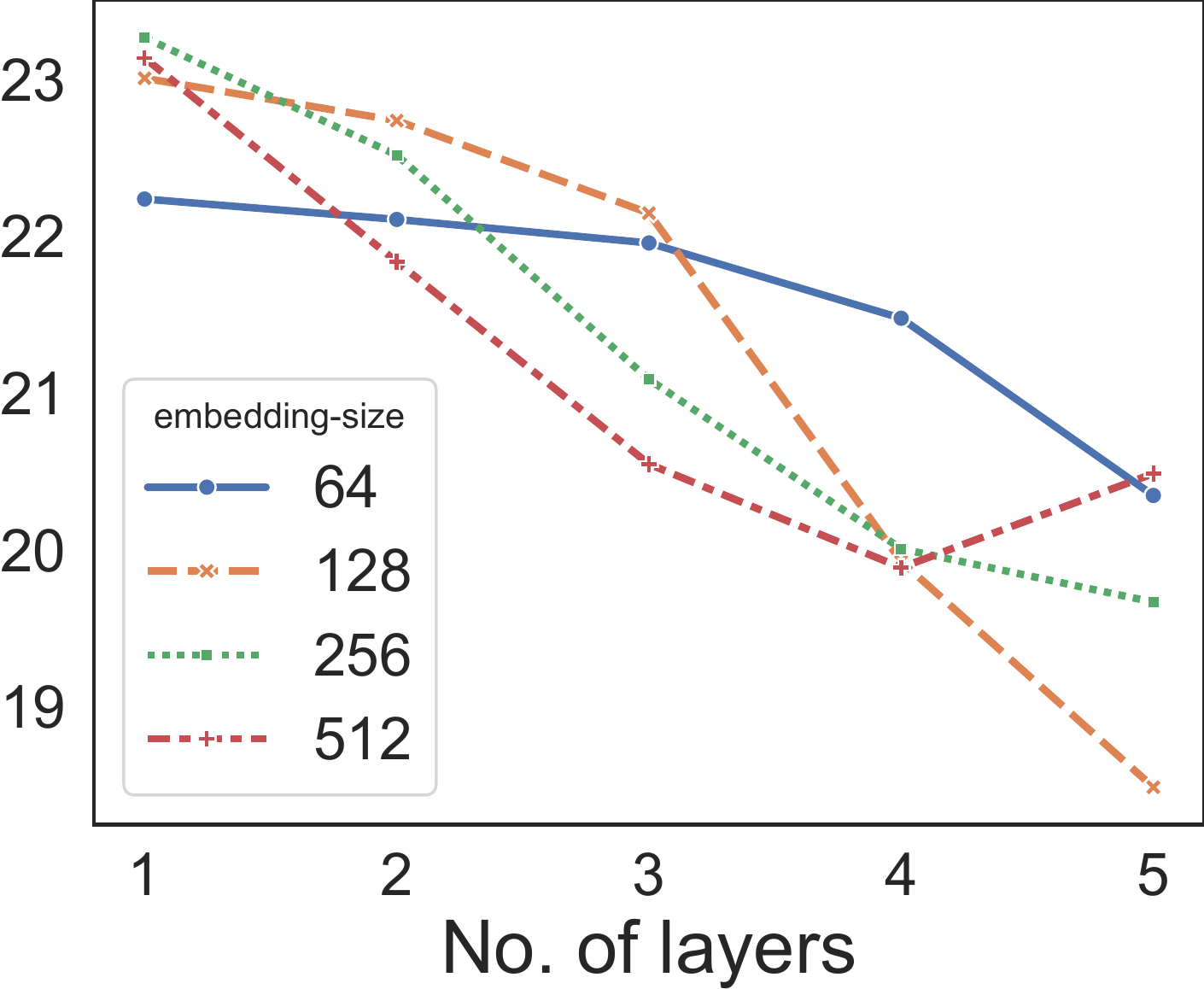}
    \label{fig:nowplaying}}
    \caption{The hyper-parameter study of GNG-ODE.}
    \label{fig:hyper}
    \vspace{-0.2cm}
\end{figure}

As shown in Figure~\ref{fig:hyper}, increasing the number of layers does not always result in better performance. For example, on Tmall and Nowplaying dataset, the optimal layer number is less than 3. The performance decreases quickly when the layer number exceeds this optimal value because of the over-smoothing problem~\cite{xu2018representation}.

\section{Conclusion}
In this paper, we design a new SBR model, GNG-ODE, to model the continuity of user preference along the time in a fully continuous manner with Neural ODE. GNG-ODE works upon our defined continuous-time temporal session graph. We employ the GGNN to encode the structural patterns to infer the initial latent states for all items. We further derive GNG-ODE that propagates the latent states of the items between different time steps in time. We also propose a time alignment algorithm, called \textit{t-Alignment}, to adapt the existing ODE solvers onto our dynamic graph setting. Extensive experiments on three real-world datasets demonstrate the effectiveness of GNG-ODE. Moreover, the ablation study and analysis verify the efficacy of those components in GNG-ODE. In conclusion, GNG-ODE is a novel model to solve the SBR problem with temporal information.

\noindent\textbf{Acknowledgement.} This work was supported by National Key Research and Development Program of China under Grant No. 2018AAA0101902, and NSFC under Grant No. 61532001.







\bibliographystyle{ACM-Reference-Format}
\balance
\bibliography{sample-base}

\appendix

\end{document}